%% file: paper.tex
\documentclass[11pt]{article}

\usepackage{acl}

\usepackage{times}
\usepackage{latexsym}
\usepackage[T1]{fontenc}
\usepackage[utf8]{inputenc}

\usepackage{microtype}
\usepackage{inconsolata}

\usepackage{amsmath, amssymb, amsthm}
\usepackage{mathtools}

\usepackage{booktabs}
\usepackage{multirow}
\usepackage{array}
\usepackage{makecell}

\usepackage{graphicx}
\usepackage{subcaption}

\usepackage{algorithm}
\usepackage{algpseudocode}

\usepackage{xcolor}
\usepackage{url}
\usepackage{hyperref}
\usepackage{xspace}
\usepackage{enumitem}

\usepackage{booktabs}
\usepackage{xspace}
\usepackage{amsmath,amssymb}

\usepackage{booktabs}
\usepackage{graphicx}
\usepackage{placeins}

\newcommand{\methodname}{TAG\xspace}

\usepackage{xcolor}
\usepackage{amsmath}        % \text{} support
\usepackage{tcolorbox}
\tcbuselibrary{breakable,skins}
\usepackage{fvextra}        % NOT fancyvrb -- supports breaklines properly

\title{Beyond Similarity: Task-Aligned Retrieval for Language Models}

\author{Zhixing Sun \\
    Beijing University of Posts \\
    and Telecommunications \\
    Beijing, China \\
  \texttt{3342349163@bupt.edu.cn} \\\And
  Shenghe Xu \\
    City University of Hong Kong \\
    Hong Kong SAR, China \\
    \texttt{shenghexu@gmail.com} \\\And
    Tao Li \\
  City University of Hong Kong \\
  Hong Kong SAR, China \\
  \texttt{li.tao@cityu.edu.hk}
    }

\begin{document}
\maketitle

\begin{abstract}
Retrieval-augmented generation (RAG) ranks passages by semantic similarity to the input, implicitly assuming that semantic similarity is a reliable indication of applicability in downstream tasks. This assumption breaks down when task success depends not on topical relevance but on applying the correct rules, constraints, or procedural guidance. In such settings, the most useful context may be the rule triggered by the input rather than the most semantically similar passage. We propose Task-Aligned Retrieval (TAG), a retrieval framework that replaces similarity-based retrieval with applicability-based rule selection. TAG transforms source documents into traceable condition-action rules, identifies which rules apply to a given input through pairwise LLM judgments, and generates the output conditioned only on the selected actions. We empirically observe that across Wikipedia NPOV rewriting, HumanEval with PEP~8 compliance, and NBA transaction reasoning on RuleArena, TAG consistently outperforms standard RAG, with the largest gains in high-mismatch settings (up to 12.2\%) while reducing retrieved context by up to 93\%. These results suggest that, in rule- and instruction-governed tasks, retrieval should optimize for applicability rather than for semantic similarity alone.
\end{abstract}

% =============================================================================
\section{Introduction}\label{sec:intro}
Retrieval-augmented generation (RAG) improves language models by conditioning generation on retrieved external context rather than relying solely on parametric memory \citep{lewis2020retrieval}. In standard RAG pipelines, retrieval is driven by semantic similarity between the input and indexed passages, resting on an implicit assumption that the context most semantically similar to the input is also the most useful for the downstream task. This is a premise that works well in fact-seeking settings, where topical relevance often correlates with evidential usefulness. 

However, this assumption fails in tasks where task completion pertains not to retrieving semantically related evidence, but to applying the correct rules, constraints, or procedural guidance. In such settings, the retrieval target is not the passage with the greatest topical overlap, but the rule applicable to the input. Revising a Wikipedia sentence, checking code against a style guide, or evaluating the legality of a transaction all require identifying the governing rule rather than the most topically similar text.

We formalize this failure mode as \emph{retrieval objective mismatch}: a divergence between the criterion for retrieving context and the criterion that determines the utility of retrieved context. In settings governed by instructional documents, applicability may depend on abstract conditions, decision boundaries, or dispersed procedural logic that semantic similarity fails to capture. As a result, retrieving the most semantically similar chunk can systematically miss the context that should actually drive generation.

%Figure~\ref{fig:motivation} illustrates this mismatch using Wikipedia's Neutral Point of View (NPOV) policy. Given an input sentence describing Hamas as a ``militant organization,'' a similarity-based retriever surfaces chunks discussing political labeling in general because they share topical vocabulary with the input. However, the governing rule is the policy requiring disputed characterizations to be attributed rather than asserted in Wikipedia’s authoritative voice. The applicable rule shares limited surface overlap with the input and is therefore ranked low by similarity-based retrieval, despite being the context that should determine the correct revision.

\begin{figure*}[t]
  \centering
  \includegraphics[width=\textwidth]{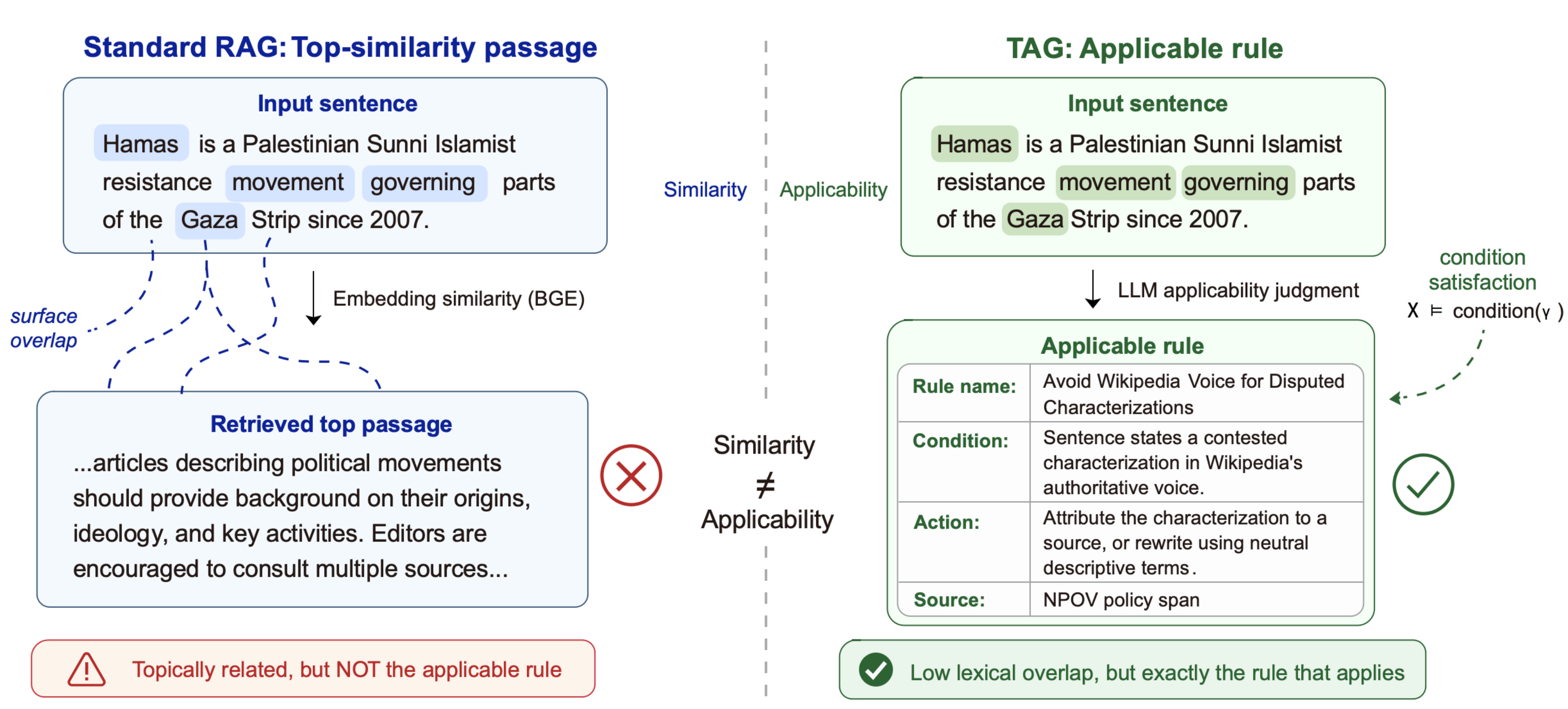}
  \caption{The retrieval objective mismatch. The top chunk retrieved by similarity discusses how to describe political movements in general, whereas the applicable rule concerns the assertion of disputed characterisations in the authoritative voice of Wikipedia. The applicable rule has low semantic similarity with the input.}
  \label{fig:motivation}
\end{figure*}

Figure~\ref{fig:motivation} illustrates this mismatch, where the LLM is tasked with revising a Wikipedia sentence to comply with the Neutral Point of View (NPOV) policy. Given an input sentence describing Hamas as a ``militant organization'', a similarity-based retriever retrieves politically related chunks due to topical overlap. However, the rule in the policy that actually applies to this sentence concerns attribution of disputed characterizations rather than how political entities should be described in general. Because this applicable rule shares little lexical overlap with the input, it is ranked low by similarity-based retrieval even though it is the context that should determine the correct revision.

\paragraph{Our framework.}
To address the retrieval objective mismatch, we propose \methodname, a task-aligned retrieval framework that alters both the workflow and the objective of retrieval. Instead of retrieving free-form text chunks by semantic similarity, \methodname first transforms the source document into atomic condition-action rules through an LLM-driven extraction pipeline, with each rule linked to a traceable source context. A pairwise applicability matcher then determines, for each rule independently, whether the task input satisfies its condition.
Finally, only the actions of the matched rules are passed to a domain-neutral executor for generation. This decomposition makes retrieval decisions explicit, executable, and traceable to the matched rules that drive generation. This shift from similarity to applicability is most helpful when the applicable rule shares little semantic similarity (measured by cosine similarity) with the input, as is often the case in regulatory compliance, and semantic similarity poorly predicts which rule applies (\S\ref{sec:discussion}).
%The key idea is to shift from semantic similarity to applicability matching when handling SOME TASKS in RAG (the task settings). 

% Our claim is specific: when source documents encode useful content as conditional rules, the retrieval objective should shift from semantic similarity to condition satisfaction. The key shift is in the retrieval objective itself, not merely the ranking procedure.

\paragraph{Contributions.}
Our contributions are threefold. First, we identify \emph{retrieval objective mismatch}, a structural failure mode of similarity-based RAG in which semantic similarity is misaligned with downstream utility. Second, we introduce \methodname, a task-aligned retrieval framework that retrieves applicability-matched atomic rules rather than semantically similar text chunks. Third, across Wikipedia NPOV rewriting, PEP~8-constrained code generation, and NBA transaction reasoning, we characterize when task-aligned retrieval helps: gains are largest under substantial retrieval mismatch, smaller when similarity and applicability already align, and limited when downstream reasoning requires complex multi-step rule interaction.

\section{Related Work}

\paragraph{Similarity-based and reranked RAG.}
Standard RAG retrieves semantically similar passages from the chunked source 
documents using dense retrieval or reranking \citep{lewis2020retrieval,
karpukhin2020dense,nogueira2019passage,glass2022rerank}. Subsequent work 
improves this pipeline through query rewriting, adaptive retrieval, 
iterative reasoning, and graph-based retrieval \citep{ma2023query,
jeong2024adaptiverag,trivedi2023interleaving,edge2024from,jiang2023active}. 
These methods improve retrieval quality while preserving semantic 
similarity as the governing retrieval relation.

\paragraph{Structured and rule-guided RAG.}
Several methods modify the retrieval unit rather than the ranking relation. 
GraphRAG \citep{edge2024from} retrieves graph neighborhoods rather than flat passages, and other variants enrich chunks with higher-level structure. 
RuleRAG uses externally provided rules as in-context guidance 
\citep{chen2024rulerag}, while RuAG mines rules from training data 
\citep{ruag2024}. \methodname differs by automatically extracting executable 
rules from the source document itself, and retrieving them based on 
applicability rather than similarity.

\paragraph{LLM-based reranking and alignment.}
\methodname is mechanically related to LLM-based rerankers and relevance 
judges \citep{nogueira2019passage,glass2022rerank,wang2024rear}, but differs 
in the retrieval predicate: instead of asking whether a passage is broadly 
relevant, \methodname evaluates whether the input satisfies a candidate 
rule's condition. Related systems such as Self-RAG \citep{asai2024selfrag}, 
AlignRAG \citep{wei2025alignrag}, chain-of-retrieval \citep{wang2025corag}, 
and DynamicRAG \citep{sun2025dynamicrag} refine or adapt retrieval during 
or after generation. \methodname instead changes the retrieval relation 
itself.

% \begin{table}[t]
%   \centering
%   \small
%   \begin{tabular}{@{}lll@{}}
%     \toprule
%     \textbf{Aspect} & \textbf{Standard RAG} & \textbf{\methodname} \\
%     \midrule
%     Knowledge unit & Text chunk & Atomic rule \\
%     Retrieval & Similarity & Applicability \\
%     Segmentation & Chunking & Auto extraction \\
%     \bottomrule
%   \end{tabular}
%   \caption{Structural comparison between Standard RAG and \methodname.}
%   \label{tab:rag-comparison}
% \end{table}
% ==========================================================================
\section{Methodology}
\label{sec:method}

\subsection{Problem Formulation}
\label{sec:method:formulation}

Let $D$ denote a source document and $T$ a downstream task whose correct outputs are determined by the rules and constraints in $D$. Standard RAG \citep{lewis2020retrieval} partitions $D$ into $N$ passages $\{c_1,\dots,c_N\}$ and retrieves the top-$k$ passages most similar to input $x$:
\begin{equation}
\mathcal{C}_{\mathrm{RAG}}(x) \;=\; \operatorname*{Top\text{-}}k_{c_i \in D}\;
\mathrm{sim}\bigl(\mathrm{emb}(x),\, \mathrm{emb}(c_i)\bigr)
\label{eq:standard-rag}
\end{equation}
% \begin{equation}
%   \mathcal{C}_{\text{RAG}}(x) = \operatorname*{Top-}k_{i=1,\dots,N}
%     \bigl\{\, \mathrm{sim}\bigl(\mathrm{emb}(x), \mathrm{emb}(c_i)\bigr) \,\bigr\},
%   \label{eq:standard-rag}
% \end{equation}
where $\mathrm{emb}(\cdot)$ is a dense encoder \citep{xiao2023cpack} and $\mathrm{sim}(\cdot,\cdot)$ is cosine similarity.

\paragraph{The retrieval objective mismatch.}
Eq.~\ref{eq:standard-rag} optimizes semantic similarity, whereas in tasks where success depends on applying the correct rules or constraints, the relevant retrieval relation is applicability. Similarity measures representational proximity, whereas applicability asks whether a rule’s condition is satisfied by the task input. Formally, a conditional rule $r$ applies to input $x$ if $\mathrm{cond}(r)$ holds for $x$, where $\mathrm{cond}(r)$ denotes the condition of $r$.
The idealized retrieval target is therefore
% Equation~\ref{eq:standard-rag} optimizes similarity, whereas for instructional documents the relevant relation is applicability. Similarity compares surface representations, while applicability compares structural roles:
% a conditional rule $r$ applies to an input $x$ if $\mathrm{cond}(r)$ holds for $x$,
% where $\mathrm{cond}(r)$ denotes the condition of $r$.
% The idealised retrieval target is therefore
\begin{equation}
\mathcal{R}^{\star}(x;D)
=
\{\, r \in \mathcal{R}(D)
\mid \mathrm{cond}(r) \text{ holds for } x \,\}.
\label{eq:ideal-retrieval}
\end{equation}
Since exact applicability is not directly observable at inference time, we approximate $\mathcal{R}^{\star}$ with a pairwise LLM matcher $\mathcal{R}_{\mathrm{matched}}(x)$ (Eq.~\ref{eq:match}).

\begin{figure*}[t]
  \centering
  \includegraphics[width=\linewidth]{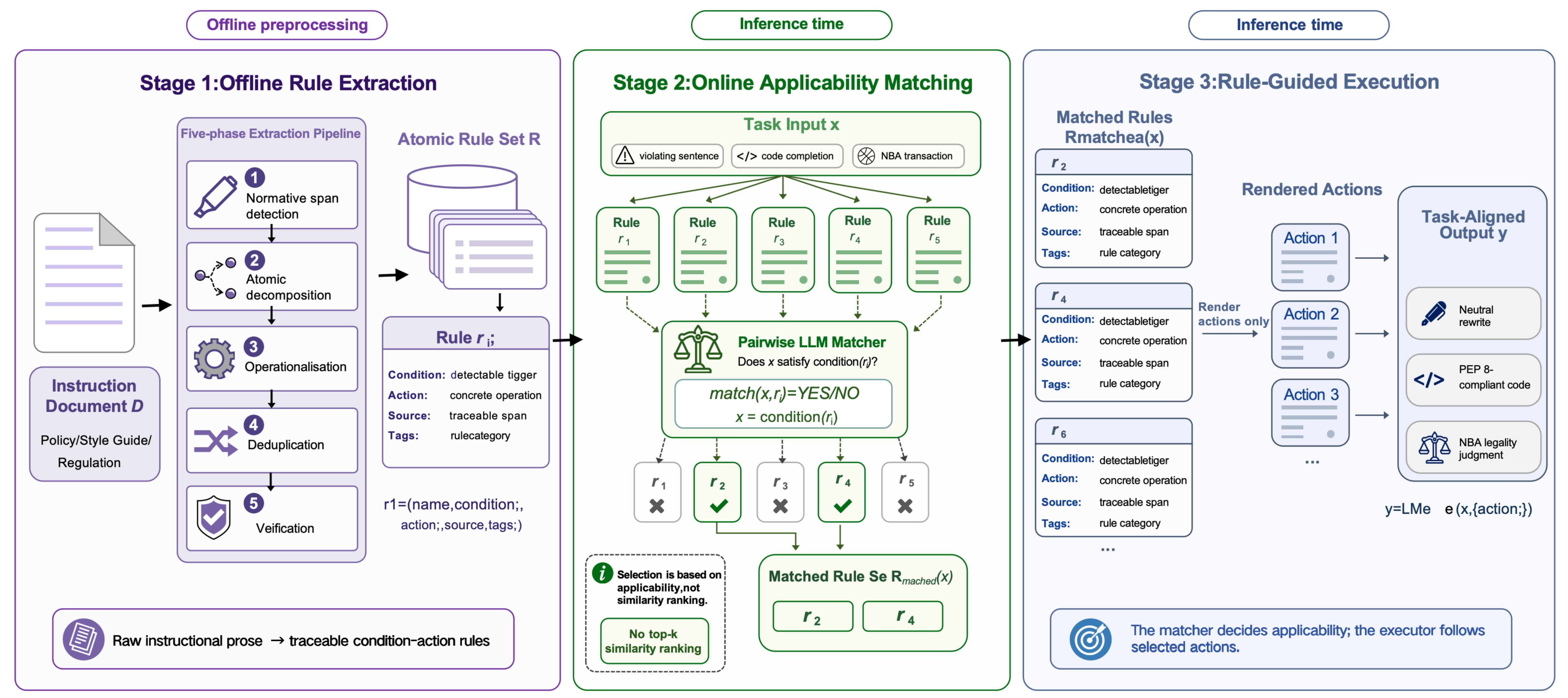}
  \caption{Overview of \methodname. Document D is processed offline into atomic rules (\S\ref{sec:method:extraction}); at inference, each rule is judged for applicability by a pairwise LLM matcher (\S\ref{sec:method:scpair}), and matched actions condition execution (\S\ref{sec:method:execution}).}
  \label{fig:pipeline}
\end{figure*}

\subsection{Framework Overview}
\label{sec:method:overview}
% \methodname instantiates three design principles through a three-stage pipeline (Figure~\ref{fig:pipeline}). P1 (atomic units): retrieval operates on atomic rules rather than character-bounded chunks. P2 (applicability relation): selection optimises condition satisfaction rather than similarity. P3 (separation of roles): the matcher selects rules; the executor follows only their actions. 

% Stage~1 (offline) converts $D$ into a rule set
\methodname follows three design principles, implemented through a three-stage pipeline (Fig.~\ref{fig:pipeline}): (P1) retrieval operates over atomic rules; (P2) selection is based on applicability; and (P3) matcher and executor roles are separated, so the matcher selects rules while the executor follows only their actions.

Stage~1 (offline) converts $D$ into a rule set

\begin{equation}
  r_i = \bigl(\, \mathrm{name}_i,\, \mathrm{condition}_i,\, \mathrm{action}_i,\,
                  \mathrm{source}_i,\, \mathrm{tags}_i \,\bigr),
  \label{eq:rule-tuple}
\end{equation}
% where $\mathrm{condition}_i$ specifies a detectable state, $\mathrm{action}_i$ a concrete operation, $\mathrm{source}_i$ a verbatim span for traceability, and $\mathrm{tags}_i$ supports deduplication.
% Stage~2 (online) evaluates each rule independently through a single-shot
% pairwise applicability judgment:
where $\mathrm{condition}_i$ specifies a judgeable condition, $\mathrm{action}_i$ a concrete operation, $\mathrm{source}_i$ a verbatim source span for traceability, and $\mathrm{tags}_i$ denotes lightweight metadata for deduplication.

Stage~2 (online) evaluates each rule independently through pairwise applicability judgment:
\begin{equation}
  \mathrm{match}(x, r_i) =
    f_{\mathrm{judge}}(x, r_i),
  \label{eq:match}
\end{equation}
where $f_{\mathrm{judge}}$ returns YES or NO. The matched set
$\mathcal{R}_{\mathrm{matched}}(x)=\{r_i : \mathrm{match}(x,r_i)=\mathrm{YES}\}$
replaces similarity-based top-$k$ retrieval.

Stage~3 prompts the executor with input $x$ and only the actions of matched rules.
% where $f_{\mathrm{judge}}$ returns YES or NO. The matched set
% $\mathcal{R}_{\mathrm{matched}}(x) =
% \{ r_i : \mathrm{match}(x, r_i) = \mathrm{YES} \}$
% replaces top-$k$ retrieval.

% Stage~3 prompts the executor with $x$ and only the actions of matched rules:
% \begin{equation}
%   y = \mathrm{LLM}_{\mathrm{exec}}\bigl(\, x,\,
%         \{\, \mathrm{action}_i : r_i \in \mathcal{R}_{\mathrm{matched}}(x)
%                                        \,\} \,\bigr).
%   \label{eq:execution}
% \end{equation}

\subsection{Stage 1: Five-Phase Rule Extraction}
\label{sec:method:extraction}

Stage~1 converts $D$ into $\mathcal{R}$ through five LLM-driven phases (Fig.~\ref{fig:extraction}), each targeting a distinct extraction failure mode.

\paragraph{Phase 1: Source text span detection.}
The LLM extracts instructional spans from $D$ while preserving source boundaries for traceability, yielding $\mathcal{S}=\{s_1,\ldots,s_L\}$ (prompt: App.~\ref{app:prompts:phase1}).

\paragraph{Phase 2: Atomic decomposition.}
Each $s_j$ is decomposed into atomic units containing a single condition--action pair, yielding $\mathcal{A}=\{a_1,\dots,a_M\}$.(App.~\ref{app:prompts:phase2})

\paragraph{Phase 3: Operationalization.}
Each atomic unit $a_k$ is converted into the structured rule tuple in Eq.~\ref{eq:rule-tuple}. The prompt enforces judgeable conditions, executable actions, traceable source spans, and lightweight category tags used for downstream comparison and deduplication.(App.~\ref{app:prompts:phase3})

\paragraph{Phase 4: Deduplication and conflict resolution.}
To avoid $O(M^2)$ comparison, we compare only rules sharing at least one tag. An LLM classifies each pair as duplicate (merged), overlap (linked but retained), conflict (flagged), or independent, yielding $M' \le M$ rules.(App.~\ref{app:prompts:phase4})

\paragraph{Phase 5: Verification.}
We apply three sanity checks: \textit{Faithfulness} (source spans remain traceable to $D$), \textit{Coverage} (each source span in $\mathcal{S}$ is represented by at least one rule), and \textit{Independence} (a duplicate-rate proxy based on rule-name uniqueness). The full verification procedure is provided in App.~\ref{app:prompts:phase5}.

\subsection{Stage 2: Pairwise Applicability Matching}
\label{sec:method:scpair}
\paragraph{Applicability prompt.}
Unlike standard LLM rerankers, which judge whether a passage is broadly relevant to the input, our matcher evaluates a narrower question: whether the input satisfies a candidate rule’s condition. Concretely, instead of asking ``Is this rule relevant to the input?'', we ask ``Does the input satisfy the condition of this rule?'' This reduces retrieval to a binary entailment-style judgment over a structured condition.
% \paragraph{Applicability prompt.}
% The judge is prompted with a strictly narrower predicate than standard LLM rerankers. Rather than ``Is this rule relevant to the input?'', it is asked ``Does the input satisfy the condition of this rule?''---a binary entailment-style decision over a structured condition. 

\paragraph{Pairwise rather than batched.}
We evaluate each pair $(x, r_i)$ in isolation rather than batching all $M$ rules into a single prompt, as batched judgments suffer from attention dilution \citep{liu2024lost} and inter-rule interference. For each pair, the judge produces a single YES/NO decision at temperature 0, and rules judged YES are added to $\mathcal{R}_{\mathrm{matched}}(x)$.(App.~\ref{app:prompts:matcher})

\subsection{Stage 3: Rule-Guided Execution}
\label{sec:method:execution}
Stage~3 prompts the executor with input $x$ and the actions of matched rules. Following P3, only $\mathrm{action}_i$ is exposed to the executor; conditions, source spans, and tags are withheld to preserve the matcher–executor separation and prevent the executor from implicitly re-ranking rules.
% Stage~3 prompts the executor with $x$ and the actions of matched rules (Equation~\ref{eq:execution}). Following P3, only $\mathrm{action}_i$ is given
% %, while $\mathrm{condition}_i$, $\mathrm{source}_i$, and $\mathrm{tags}_i$ are withheld so that 
% to the executor, so it follows the matcher's selection without reassessing applicability. 
% =============================================================================
\section{Experiments}
\label{sec:experiments}

\subsection{Research Questions and Experiment Design}
\label{sec:exp:rq}
Our experiments are organized around four research questions. RQ1 asks whether retrieval objective mismatch exists in practice: do applicability-based methods yield larger gains in domains where semantic similarity is poorly aligned with rule applicability? RQ2 asks whether replacing top-$k$ similarity retrieval with pairwise applicability matching improves downstream generation across diverse tasks. RQ3 asks which design component drives the gain: rule structuring (P1), applicability-based selection (P2), or action-only execution (P3). This question is addressed by the ablations in \S\ref{sec:ablation1} and \S\ref{sec:ablation2}. RQ4 asks where \methodname fails, i.e., which task characteristics limit its effectiveness.

% Our experiments are organised around four research questions. RQ1 asks whether the applicability gap exists, that is, whether applicability-based retrieval yields larger gains in domains where semantic similarity is less aligned with rule applicability.
% RQ2 asks whether applicability-based retrieval improves downstream generation, that is, whether substituting pairwise applicability matching for top-k similarity yields better outputs across diverse tasks. RQ3 asks which component contributes to the gain: rule structuring (P1), the applicability relation (P2), or action-only execution (P3). This question is addressed by the ablations in \S\ref{sec:exp:ablation}. RQ4 asks when \methodname fails, that is, for which kinds of tasks the framework is most and least beneficial.

\subsection{Setup}
\label{sec:exp:setup}

\paragraph{Domains.}
We evaluate on three domains chosen to vary in the degree of mismatch between semantic similarity and rule applicability. \textbf{Wikipedia NPOV} (text rewriting, $107$ cases from the NPOV Noticeboard 2020--2024) is a high-gap setting: given a sentence flagged for a neutrality violation, the model must rewrite it to remove the violation while preserving factual content, and violations often share little lexical overlap with the governing rules. \textbf{HumanEval~$\times$~PEP~8} \citep{chen2021codex} is a code-generation and style-compliance task built from $164$ HumanEval programming problems. Given a function signature and docstring, the model must generate a function body that both passes the official unit tests and conforms to PEP~8. \textbf{RuleArena-NBA} \citep{zhou2025rulearena} ($156$ problems stratified into L0/L1/L2 by transaction count, see Appendix \ref{app:setup} for detail) is a high-conditionality multi-step setting: each problem asks whether one or more proposed transactions are permitted under the NBA Collective Bargaining Agreement, and asks the model to identify the offending transaction and responsible team. Dataset details are in Appendix~\ref{app:setup}.

\paragraph{Metrics.}
For NPOV we report a binary Violation Fix Rate (VFR) plus four 1--5 auxiliary scores (Remediation, Preservation, Tone, Fluency). For code we report Pass@1 and pylint (headline style metric). For RuleArena-NBA we follow the benchmark's strict accuracy, which requires the True/False answer, the offending transaction, and the responsible team to all be correct. Scoring details and evaluator prompts are in Appendix~\ref{app:metrics}.

\paragraph{Models and baselines.}
DeepSeek-V4 serves as both the executor and the pairwise matcher; we also report results with Qwen3-30B-A3B-Instruct-2507 as executor. We compare against \textbf{M0} (no retrieval), \textbf{M1} (all extracted rules in the prompt), 
%\textbf{M2} (Standard RAG over 500-character chunks with 100-character overlap, indexed by BGE \citep{xiao2023cpack}, top-$k$ for $k\!\in\!\{5,10,15,20\}$), 
\textbf{M2} (Standard RAG over 500-character chunks with 100-character overlap, indexed with BGE \citep{xiao2023cpack}, with top-$k$ for $k\in\{5,10,15,20\}$). \textbf{M3 (\methodname, Ours)} performs single-shot pairwise applicability judgment at temperature 0 for each (input, rule) pair and conditions the executor on the actions of matched rules.
% -----------------------------------------------------------------------------

% -----------------------------------------------------------------------------
\subsection{Experiment 1: Wikipedia NPOV}
\label{sec:exp:wiki}

\paragraph{Setup.}
We collected $107$ NPOV violation cases from the Noticeboard, where editors discuss article-level disputes about whether content satisfies the Neutral Point of View policy. Each case pairs a flagged sentence with an editor-provided description of the neutrality issue. We process the NPOV policy ($\sim$15K words) into $M{=}119$ atomic rules; faithfulness, coverage, and independence all exceed $0.94$. Full curation criteria, extraction statistics, and the trivial-rewrite filter are in Appendix~\ref{app:setup}.

\begin{table*}[t]
  \centering
  \small
  \begin{tabular}{@{}llrrrrrrr@{}}
    \toprule
    Executor & Method
        & \makecell{\#Rules}
        & VFR (\%)
        & Rem & Pres & Tone & Flu
        & Avg \\
    \midrule
    \multirow{7}{*}{\makecell[l]{DeepSeek-V4}}
      & No Retrieval (M0)             & 0.0   & 32.7 & 2.33 & \textbf{4.70} & 3.44 & \textbf{4.58} & 3.76 \\
      & All Rules (M1)                & 131.0 & 63.5 & 3.51 & 4.24 & \textbf{3.96} & 4.21 & 3.98 \\
      & Std RAG, top-5                & 5.0   & 55.1 & 3.34 & 4.36 & 3.78 & 4.25 & 3.93 \\
      & Std RAG, top-10               & 10.0  & 59.8 & 3.34 & 4.38 & 3.76 & 4.08 & 3.89 \\
      & Std RAG, top-15               & 15.0  & 59.8 & 3.39 & 4.40 & 3.79 & 4.30 & 3.97 \\
      & Std RAG, top-20               & 20.0  & 60.8 & 3.48 & 4.44 & 3.84 & 4.29 & \textbf{4.01} \\
      & \methodname (Ours)            & 21.0  & \textbf{68.2} & \textbf{3.55} & 4.13 & 3.94 & 4.21 & 3.96 \\
    \midrule
    \multirow{7}{*}{\makecell[l]{Qwen3-30B-A3B\\-Instruct-2507}}
      & No Retrieval (M0)             & 0     &  6.5 & 1.25 & \textbf{4.94} & 3.13 & \textbf{5.00} & 3.58 \\
      & All Rules (M1)                & 131   & 62.6 & 3.33 & 3.99 & 3.92 & 4.33 & \textbf{3.89} \\
      & Std RAG, top-5                & 5     & 57.0 & 3.28 & 3.76 & \textbf{3.97} & 4.35 & 3.84 \\
      & Std RAG, top-10               & 10    & 55.1 & 3.21 & 3.79 & 3.92 & 4.30 & 3.81 \\
      & Std RAG, top-15               & 15    & 58.9 & 3.31 & 3.79 & 3.96 & 4.35 & 3.85 \\
      & Std RAG, top-20               & 20    & 53.3 & 3.13 & 3.80 & 3.83 & 4.27 & 3.76 \\
      & \methodname (Ours)            & 32.6  & \textbf{64.5} & \textbf{3.45} & 3.70 & 3.90 & 3.79 & 3.71 \\
    \bottomrule
  \end{tabular}
  \caption{Results on Wikipedia NPOV rewriting ($n=107$). VFR is the binary Violation Fix Rate; Rem/Pres/Tone/Flu are 1--5 auxiliary scores; Avg is their mean. \#Rules reports the average number of rules (or chunks, for Standard RAG) provided to the executor.}
  \label{tab:wiki-main}
\end{table*}
% \paragraph{Results.}
% Table~\ref{tab:wiki-main} reports the main results. On both executors, \methodname attains the top VFR among all retrieval configurations: $68.2\%$ on DeepSeek-V4 and $64.5\%$ on Qwen3-30B-A3B-Instruct-2507, exceeding the strongest Standard RAG variant by $7.4$ and $5.6$ absolute points respectively. Standard RAG itself shows no monotonic improvement as $k$ grows from $5$ to $20$ ($55.1$--$60.8\%$ on DeepSeek-V4, non-monotonic on Qwen3 with the largest $k$ yielding the lowest VFR), indicating that enlarging a similarity-driven candidate pool does not reliably surface the governing rules. The comparison to the All-Rules baseline is also informative: M1 is the strongest baseline on both executors ($63.5\%$ and $62.6\%$), yet \methodname improves on it while reducing the average rule count from $131.0$ to $21.0$ and $32.6$, a $6.2\times$ and $4.0\times$ context reduction. The gain is therefore not an artifact of exposing the executor to more rule-structured content. On the auxiliary 1--5 scores, \methodname leads on Remediation in both panels ($3.55$ and $3.45$) at a small cost to Preservation and Fluency, the expected signature of a method that rewrites more aggressively to resolve more violations. This is the predicted pattern: NPOV is the high-gap domain, and the gain manifests as a large headline improvement in VFR rather than a marginal one.

\paragraph{Results and analysis.}
Table~\ref{tab:wiki-main} reports the main results. \methodname achieves the highest Violation Fix Rate (VFR) on both executors, reaching $68.2\%$ on DeepSeek-V4 and $64.5\%$ on Qwen3-30B-A3B-Instruct-2507, outperforming the strongest Standard RAG baseline by $7.4$ and $5.6$ absolute points respectively.
Standard RAG shows no consistent improvement as retrieval depth increases for both LLMs, with the largest candidate pool yielding the worst result for Qwen. This suggests that simply enlarging a similarity-driven retrieval pool does not reliably find the governing policy rules.
The All-Rules baseline isolates the effect of rule structuring without selection. Although it is the strongest non-\methodname baseline on both LLMs ($63.5\%$ and $62.6\%$), \methodname still improves upon it while reducing average context size from $131.0$ rules to $21.0$ and $32.6$, corresponding to $6.2\times$ and $4.0\times$ reductions. This shows that the improvement cannot be explained by rule structuring alone: selecting the applicable subset is more effective than conditioning the executor on the full rule set.
On the auxiliary metrics, \methodname achieves the highest Remediation score in both settings ($3.55$ and $3.45$), with modest decreases in Preservation and Fluency, suggesting that TAG prioritizes remediation when neutrality preservation trades off against fluency.

We validate the LLM judge with human annotation on $n=40$ cases by 
two independent raters; inter-annotator agreement is substantial 
($\kappa \in [0.68, 0.75]$) and the ranking M3 $>$ M1 $>$ M2 is 
preserved across all metrics. See Appendix~\ref{app:human_eval}.

% -----------------------------------------------------------------------------
\subsection{Experiment 2: HumanEval with PEP~8 Compliance}
\label{sec:exp:code}

%PEP~8 is included as a deliberate boundary test. Many style rules are directly reflected in surface code patterns: an indentation rule is applicable to code with incorrect indentation, and a naming rule is applicable to identifiers that violate naming conventions. Similarity and applicability are therefore much more closely aligned than in NPOV, reducing the room for improvement from applicability-based retrieval.

\paragraph{Setup.}
HumanEval \citep{chen2021codex} provides $164$ function-completion tasks. PEP~8 ($\sim$8K words) is processed into $127$ atomic rules. Pass@1 is computed following the official HumanEval protocol; pylint configurations
are detailed in Appendix~\ref{app:metrics}..

\begin{table*}[t]
  \centering
  \small
  \begin{tabular}{@{}llrrr@{}}
    \toprule
    Executor & Method
        & \makecell{\#Units}
        & Pass@1 (\%)
        & Pylint \\
    \midrule
    \multirow{7}{*}{\makecell[l]{DeepSeek-V4}}
      & No Retrieval (M0)             & 0     & \textbf{98.8} & 9.12 \\
      & All Rules (M1)                & 127   & \textbf{98.8} & 9.32 \\
      & Std RAG, top-5                & 5     & 97.6          & 9.13 \\
      & Std RAG, top-10               & 10    & 98.4          & 9.21 \\
      & Std RAG, top-15               & 15    & 98.2          & 9.20 \\
      & Std RAG, top-20               & 20    & \textbf{98.8} & 9.11 \\
      & \methodname (Ours)            & 22.0  & 98.1          & \textbf{9.36} \\
    \midrule
    \multirow{7}{*}{\makecell[l]{Qwen3-30B-A3B\\-Instruct-2507}}
      & No Retrieval (M0)             & 0     & \textbf{93.3} & 8.26 \\
      & All Rules (M1)                & 127   & 90.2          & 9.23 \\
      & Std RAG, top-5                & 5     & 87.8          & 8.63 \\
      & Std RAG, top-10               & 10    & 88.4          & 8.80 \\
      & Std RAG, top-15               & 15    & 88.4          & 8.70 \\
      & Std RAG, top-20               & 20    & 89.0          & 8.85 \\
      & \methodname (Ours)            & 36.1  & 90.2          & \textbf{9.27} \\
    \bottomrule
  \end{tabular}
  \caption{Results on HumanEval with PEP~8 compliance ($n=164$). Pylint
  serves as the headline style metric; Radon and Bandit are reported as
  secondary indicators.}
  \label{tab:code-main}
\end{table*}

\paragraph{Results and analysis.}
Table~\ref{tab:code-main} reports the results. \methodname achieves the highest Pylint score on both LLMs, reaching $9.36$ on DeepSeek-V4 and $9.27$ on Qwen3-30B-A3B-Instruct-2507. Compared to Wikipedia NPOV, the gains over the strongest Standard RAG baseline are smaller ($+0.15$ on DeepSeek-V4 and $+0.42$ on Qwen3), indicating less room for applicability-based retrieval to improve over similarity-based retrieval in this setting.
Pass@1 provides a complementary robustness check. On DeepSeek-V4, functional correctness remains essentially unchanged across methods ($97.6$--$98.8\%$), indicating that stronger style compliance does not come at a meaningful cost to task completion. On Qwen3, No Retrieval achieves the highest Pass@1 ($93.3\%$), while retrieval-augmented variants cluster near $90\%$, suggesting a modest functionality trade-off in this setting.
The All-Rules baseline is also highly competitive, reaching Pylint scores of $9.32$ and $9.23$, close to \methodname. This indicates that structured rule conditioning already captures much of the available gain in this domain.

\subsection{Experiment 3: RuleArena-NBA (High Conditionality, Multi-Step)}
\label{sec:exp:nba}

\paragraph{Setup.}
We use the NBA subset of RuleArena \citep{zhou2025rulearena} ($156$ problems across L0/L1/L2). The NBA CBA ($\sim$60K words) is processed into $195$ atomic rules. See Appendix~\ref{app:setup} for the per-level breakdown and rule coverage.

\begin{table*}[t]
  \centering
  \small
  \begin{tabular}{@{}llrrrrr@{}}
    \toprule
    Executor & Method
        & \makecell{\#Units}
        & \makecell{Strict (\%)} & L0 (\%) & L1 (\%) & L2 (\%) \\
    \midrule
    \multirow{7}{*}{\makecell[l]{DeepSeek-V4}}
      & No Retrieval (M0)             & 0     & 42.9 & 54.1 & 43.5 & 15.4 \\
      & All Rules (M1)                & 195   & 39.7 & 50.8 & 39.1 & 15.4 \\
      & Std RAG, top-5                & 5     & 40.4 & 55.7 & 37.7 & 11.5 \\
      & Std RAG, top-10               & 10    & 43.6 & 49.2 & \textbf{46.4} & \textbf{23.1} \\
      & Std RAG, top-15               & 15    & 34.6 & 47.5 & 29.0 & 19.2 \\
      & Std RAG, top-20               & 20    & 42.9 & 55.7 & 42.0 & 15.4 \\
      & \methodname (Ours)            & 20.8  & \textbf{44.2} & \textbf{65.6} & 36.2 & 15.4 \\
    \midrule
    \multirow{7}{*}{\makecell[l]{Qwen3-30B-A3B\\-Instruct-2507}}
      & No Retrieval (M0)             & 0     & 30.1 & 37.7 & 29.0 & 15.4 \\
      & All Rules (M1)                & 195   & 34.0 & 45.9 & 29.0 & \textbf{19.2} \\
      & Std RAG, top-5                & 5     & 32.1 & 42.6 & 29.0 & 15.4 \\
      & Std RAG, top-10               & 10    & 31.4 & 42.6 & 29.0 & 11.5 \\
      & Std RAG, top-15               & 15    & 29.5 & 39.3 & 24.6 & \textbf{19.2} \\
      & Std RAG, top-20               & 20    & 32.7 & \textbf{47.5} & 26.1 & 15.4 \\
      & \methodname (Ours)            & 14.0  & \textbf{35.9} & \textbf{47.5} & \textbf{31.9} & \textbf{19.2} \\
    \bottomrule
  \end{tabular}
  \caption{Results on RuleArena-NBA ($n=156$). \textit{\#Units} reports the average number of retrieval units supplied to the executor: 500-character chunks for Standard RAG, and atomic rules for All Rules and \methodname; M0 retrieves nothing. Strict accuracy requires the True/False answer, the offending transaction, and the responsible team to all be correct. L0/L1/L2 are complexity-stratified breakdowns by single, two, and three-or-more interacting transactions.}
  \label{tab:nba-main}
\end{table*}

\paragraph{Results and analysis.}
Table~\ref{tab:nba-main} shows that RuleArena-NBA remains challenging under the strict metric, with \methodname achieving the highest overall strict accuracy on both LLMs: $44.2\%$ on DeepSeek-V4 and $35.9\%$ on Qwen3-30B-A3B-Instruct-2507.
The level-stratified breakdown helps explain where these gains arise. On DeepSeek-V4, \methodname performs particularly strongly on the single-transaction L0 slice, reaching $65.6\%$ compared with $55.7\%$ for the strongest similarity-based baseline. On Qwen3, \methodname achieves the best or tied-best result across all three levels. This pattern suggests that applicability-based retrieval is especially effective when success depends primarily on identifying the governing rules.
Performance differences are less pronounced on the more complex L1/L2 cases, where solving the task additionally requires composing multiple interacting rules after retrieval. This indicates that RuleArena-NBA combines two distinct challenges: retrieving the relevant constraints and performing downstream multi-step rule composition. \methodname directly addresses the former, while the latter remains LLM-dependent.
\methodname is also substantially more context-efficient than the All-Rules baseline, using only $20.8$ and $14.0$ retrieved rules instead of $195$, while still achieving higher strict accuracy on both LLMs. 
Standard RAG remains non-monotonic with retrieval depth (e.g., DeepSeek-V4 drops from 43.6\% at top-10 to 34.6\% at top-15), indicating that larger similarity-based candidate pools do not reliably improve rule selection.

\subsection{Cross-Domain Analysis}
Taken together, the three domains reveal a consistent pattern: \methodname is most beneficial when semantic similarity is a weak proxy for the context that governs downstream success. In Wikipedia NPOV, where the relevant policy rules often share limited lexical overlap with the input, \methodname delivers its largest gains over similarity-based retrieval. In HumanEval with PEP~8 compliance, where many applicable rules are directly reflected in code structure, similarity-based retrieval is already comparatively effective, leaving less room for applicability-based selection to improve performance. RuleArena-NBA reveals a different limitation: while \methodname achieves the strongest overall strict accuracy, the level-stratified breakdown shows that retrieval quality is only one component of task difficulty.

A second consistent finding is that better retrieval does not require more context. Across all three domains, \methodname matches or exceeds the strongest baselines while conditioning the LLMs on substantially fewer retrieved units than the All-Rules baseline. This suggests that the gains arise from more precise rule selection rather than simply providing more structured context.

Overall, the results support the central claim of this work: in rule and instruction governed tasks, retrieval should be aligned with applicability rather than defaulting to semantic similarity.

\subsection{Ablation: Rule Extraction Phases}
\label{sec:ablation1}

To assess the contribution of each phase in the rule extraction pipeline (§\ref{sec:method:extraction}), we perform leave-one-out ablations on two domains: Wikipedia NPOV with DeepSeek-V4 as executor, and RuleArena-NBA with Qwen3-30B-A3B-Instruct-2507. In each variant, the matcher (M3-pair, temperature 0) and evaluation protocol are fixed; only the extracted rule set varies. Table~\ref{tab:ablation1} reports the results.
% To assess the contribution of each phase in the rule extraction
% pipeline (§\ref{sec:method:extraction}), we run leave-one-out
% variants on two domains: Wikipedia NPOV with DeepSeek-V4 as
% executor, and RuleArena-NBA with Qwen3-30B-A3B-Instruct-2507 as
% executor. The two settings differ in three respects: source-document
% length ($\sim$15K vs.\ $\sim$60K words), document style (prescriptive
% prose vs.\ procedural regulation), and executor scale. Running the same ablation in both settings tests whether the observed
% phase contributions are specific to one domain/executor pair or persist
% under substantially different document and model conditions. In every
% variant, the matcher (M3-pair, $T{=}0$) and the evaluation protocol
% are held fixed, so the rule set handed to the matcher is the only
% factor that varies. Table~\ref{tab:ablation1} reports both panels.
\begin{table}[t]
\centering\footnotesize
\begin{tabular}{lrrr}
\toprule
Variant & $M$ & \#picks & Metric (\%) \\
\midrule
\multicolumn{4}{l}{\textit{(a) NPOV (DeepSeek-V4)}} \\
\midrule
\textbf{Full} (5 phases)            & 119 & 22.1 & \textbf{68.2} \\
$-$ Phase 1 (span detection)        & 178 & 39.4 & 65.4 \\
$-$ Phase 2 (decomposition)         &  84 & 19.6 & 64.5 \\
$-$ Phase 3 (operational.)          & 118 & 28.4 & 63.6 \\
$-$ Phase 4 (dedup.)                & 127 & 29.7 & 65.4 \\
$-$ Phase 5 (verification)          & 127 & 24.8 & 64.5 \\
\midrule
% \multicolumn{4}{l}{\textit{\hspace{0.5em}matcher variant (same rule set)}} \\
$-$ Relevance ranker (§\ref{app:prompts:rerank-baseline}) & 119 & 34.3 & 63.6 \\
\midrule
\multicolumn{4}{l}{\textit{(b) RuleArena-NBA (Qwen3-30B)}} \\
\midrule
\textbf{Full} (5 phases)            & 195 & 14.0 & \textbf{35.9} \\
$-$ Phase 1 (span detection)        & 245 & 17.5 & 34.4 \\
$-$ Phase 2 (decomposition)  & 165 & 13.0 & 32.3 \\
$-$ Phase 3 (operational.)    & 196 & 34.0 & 29.2 \\
$-$ Phase 4 (dedup.)         & 230 & 22.0 & 31.3 \\
$-$ Phase 5 (verification)          & 205 & 14.5 & 30.2 \\
\bottomrule
\end{tabular}
\caption{Ablation on the five-phase rule extraction pipeline across
two domains. Panel (a): NPOV ($n{=}107$) with DeepSeek-V4. Panel (b):
RuleArena-NBA ($n{=}156$) with Qwen3-30B. 
%Each row removes one phase from the full pipeline. 
$M$ is the final rule-set size;
\#picks is the average number of rules retrieved.
% The final row of panel~(a) is not a phase ablation but a
% \emph{matcher variant}: it reuses the full rule set but replaces the
% applicability judgment with a relevance judgment
%(§\ref{app:prompts:rerank-baseline}).
}
\label{tab:ablation1}
\end{table}
%\paragraph{Results.}
Table~\ref{tab:ablation1} shows that Phase~3 (operationalization) is
the largest contributor in both domains, reducing performance by $4.6$
VFR points on NPOV and $6.7$ strict-accuracy points on RuleArena-NBA
when removed. The remaining phases form a second tier whose ordering
shifts with document structure: on NPOV, Phases~2 (atomic
decomposition) and 5 (verification) contribute most ($-3.7$ each),
while on NBA the second tier is led by Phase~5 ($-5.7$) and Phase~4
(deduplication, $-4.6$). Phase~1 (span detection) is never the
dominant factor in either setting.

%\paragraph{Analysis.}
The dominant effect of Phase~3 suggests that extraction alone is
insufficient: rules must also be operationalized into concrete,
judgeable conditions and executable actions. Removing this phase
increases the number of matched rules in both domains
($22.1\rightarrow28.4$ on NPOV; $14.0\rightarrow34.0$ on NBA),
indicating that vague conditions reduce matcher selectivity while
providing less actionable guidance to the executor. The primacy of
operationalization across a short policy document and a substantially
longer procedural rulebook suggests that this effect reflects a
structural property of the pipeline rather than a domain-specific
artifact. See detailed per-phase analysis in
Appendix~\ref{app:ablation-details-1}.

Table \ref{tab:ablation1} also includes a relevance ranker result (§\ref{app:prompts:rerank-baseline}). A controlled relevance-reranking baseline shows that \methodname’s gains do not come merely from using an LLM judge: replacing applicability matching with relevance matching drops performance to 63.6\% VFR, essentially identical to the All-Rules baseline (63.5\%) and 4.6 points below \methodname (68.2\%). This suggests that the key improvement comes from framing retrieval around applicability rather than generic LLM-based relevance judgment.

% \paragraph{Applicability vs.\ LLM judgment.}
% To test whether \methodname's gains stem from the applicability
% framing or merely from using an LLM judge, we evaluate a controlled
% relevance reranker: same pairwise setup and identical atomic-rule set
% as \methodname, but with the applicability predicate of
% §\ref{sec:method:scpair} (``Does the input satisfy this rule's
% condition?'') replaced by a relevance one (``Is this rule relevant?'';
% prompt in Appendix~\ref{app:prompts:rerank-baseline}). As the final row
% of Table~\ref{tab:ablation1}(a) shows, it reaches only $63.6$
% VFR---essentially the All-Rules baseline ($63.5$;
% Table~\ref{tab:wiki-main}) and $4.6$ points below \methodname
% ($68.2$). Ranking rules by relevance thus yields no gain over
% including all of them; the improvement appears only once the predicate
% shifts to applicability, confirming that the framing itself, not LLM
% judgment, drives \methodname's performance.

% \paragraph{Results.}
% Table~\ref{tab:ablation1} shows broadly consistent phase contributions across both domains. Phase~3 (operationalization) is the largest contributor, reducing performance by $6.5$ VFR points on NPOV and $6.7$ strict-accuracy points on RuleArena-NBA when removed. Phases~5 (verification), 2 (atomic decomposition), and 4 (deduplication) also contribute meaningfully, while Phase~1 (span detection) has only a minor effect in both settings.

\subsection{Ablation: Disentangling Retrieval Unit from Retrieval Relation}
\label{sec:ablation2}
\methodname changes two aspects of standard RAG at once: the retrieval
\emph{unit} (text chunk \(\to\) atomic rule) and the retrieval \emph{relation}
(BGE similarity \(\to\) LLM pairwise applicability). To disentangle these
effects, we run a $2\times2$ factorial ablation on Wikipedia NPOV, introducing
the two off-diagonal conditions: \textsc{Rules+Similarity} (BGE retrieval over
the same atomic-rule corpus used by \methodname) and
\textsc{Chunks+Applicability} (pairwise applicability matching over raw text
chunks). All four settings share the same Qwen3-30B-A3B-Instruct-2507 executor
and evaluation protocol.
Table~\ref{tab:ablation2} reports the results.
 We use Wikipedia NPOV for this ablation because
its large mismatch between semantic similarity and rule applicability makes
component-level differences easier to observe.
\begin{table}[t]
\centering\small
\begin{tabular}{llrr}
\toprule
Unit & Relation & \#picks & VFR (\%) \\
\midrule
Chunks       & Similarity (BGE)        & 20.0 & 53.3 \\
Chunks       & Applicability (pair)    & 21.4 & 59.8 \\
Atomic rules & Similarity (BGE)        & 21.0 & 54.2 \\
Atomic rules & Applicability (pair)    & 32.6 & \textbf{64.5} \\
\bottomrule
\end{tabular}
\caption{$2\times2$ ablation disentangling the retrieval unit (chunks vs.\
atomic rules) from the retrieval relation (BGE similarity vs.\ LLM pairwise
applicability) on Wikipedia NPOV ($n{=}107$, executor Qwen3-30B). 
%The
%two corner cells (Chunks~+~Similarity and Atomic~rules~+~Applicability) are
%reproduced from Table~\ref{tab:wiki-main}, while the off-diagonal cells are introduced
%by this ablation. \#picks denotes the average number of retrieved units
%passed to the executor.
}
\label{tab:ablation2}
\end{table}
Both factors contribute, but the retrieval relation contributes more. With the
unit fixed at chunks, replacing similarity with applicability raises VFR from
$53.3$ to $59.8$ (+6.5 points); with the unit fixed at atomic rules, the same
replacement raises VFR from $54.2$ to $64.5$ (+10.3). By comparison, changing
the retrieval unit yields smaller gains: +0.9 points under similarity and
+4.7 under applicability.
This rules out a simple cleaner-unit explanation: even when operating over the
same atomic-rule corpus, applicability-based retrieval substantially
outperforms similarity-based retrieval ($64.5$ vs.\ $54.2$). Rule structuring
remains complementary, but the dominant gain comes from aligning retrieval with
the downstream applicability relation rather than semantic similarity. See
detailed analysis in Appendix~\ref{app:ablation-details-2}.

\section{Conclusion}

We introduced \emph{retrieval objective mismatch}, a failure mode in rule and instruction governed tasks where semantic similarity is misaligned with downstream utility. 
%to identify the context that actually governs downstream behavior. 
To address this, we proposed \methodname, which transforms source documents into condition--action rules and retrieves context through applicability judgments rather than similarity ranking.

Across three domains, \methodname consistently outperforms standard similarity-based RAG while using substantially fewer retrieved units than all-rule conditioning, indicating that the gains arise from more precise rule selection rather than simply providing more structured context.

More broadly, our findings suggest that retrieval should be aligned with the structure of the downstream task rather than defaulting to semantic similarity. When task success depends on conditional rules, constraints, or procedural guidance, the most useful context may not be the most similar passage, but the rule applicable to the input.

\section*{Limitations}

TAG assumes that source documents can be faithfully transformed into explicit condition--action rules. Tasks whose governing knowledge is highly implicit, ambiguous, or primarily factual rather than procedural may be less suitable for this formulation. Our current pairwise applicability matching also introduces additional inference cost compared with dense retrieval, although this is partly offset by substantially reduced context passed to the executor. Our RuleArena-NBA results also suggest a second boundary: improving retrieval helps when rule selection is the primary bottleneck, but does not by itself resolve tasks requiring complex multi-step rule composition. Finally, while we evaluate across three distinct domains, broader validation on additional instruction- and policy-governed tasks would strengthen the generality of our conclusions.

% =============================================================================

% ==========================================================================
%\bibliographystyle{acl_natbib}  % ACL 模板自带
\bibliography{references}

\appendix

%\section{Five-phase rule extraction applied to a passage from the NPOV policy.}
\begin{figure*}[t]
  \centering
  \includegraphics[width=\linewidth]{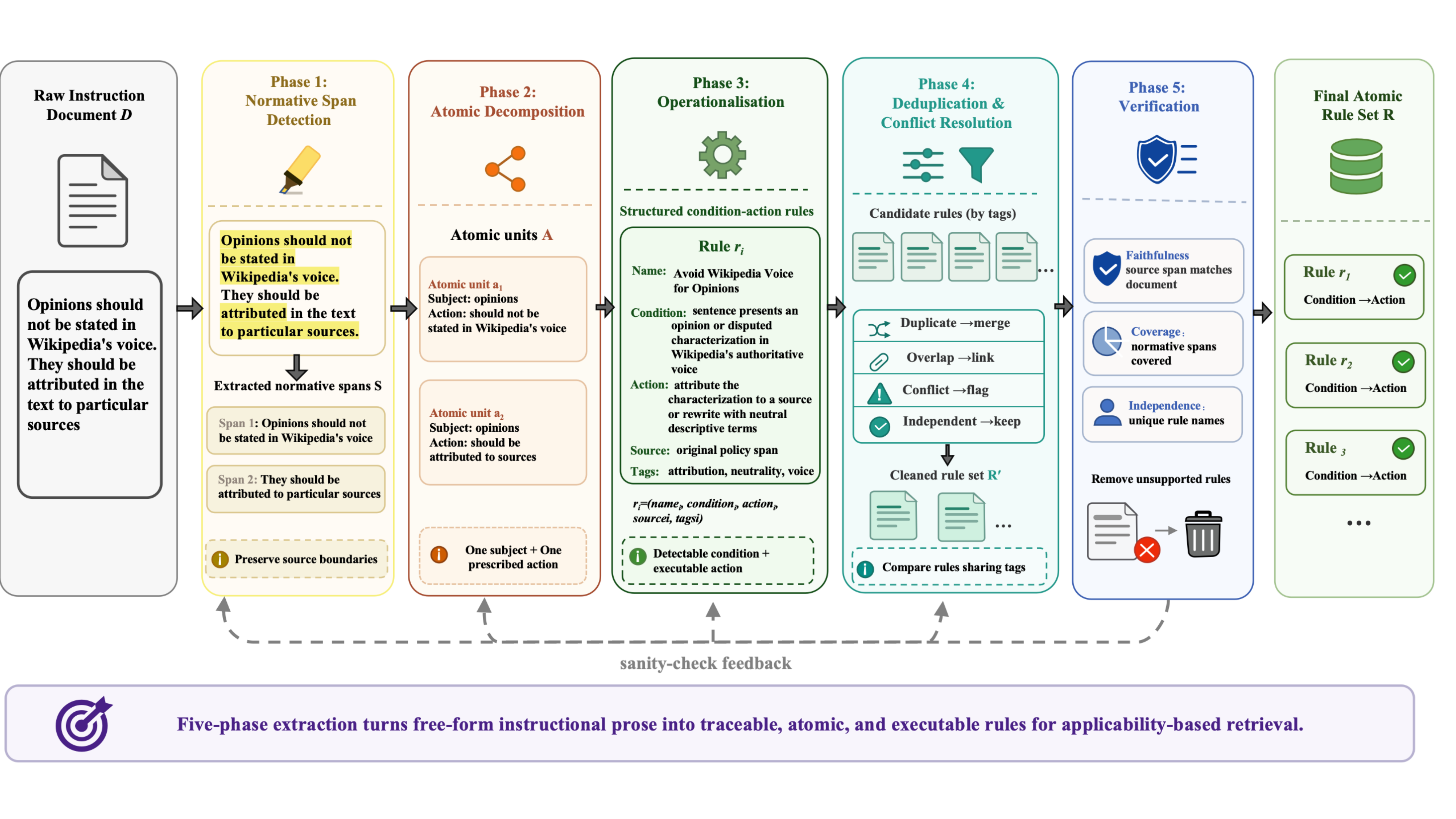}
  \caption{Five-phase rule extraction applied to a passage from the NPOV policy.}
  \label{fig:extraction}
\end{figure*}

\section{Setup Details}
\label{app:setup}

\paragraph{Wikipedia NPOV.}
Cases are drawn from the Wikipedia NPOV Noticeboard between 2020 and 2024. Each case consists of a flagged sentence, a violation description written by an editor, and a link to the source article. We retain only cases for which the violation is localised to a single sentence and the editor's description is specific enough to anchor objective scoring, yielding $107$ cases. The NPOV policy ($\sim$15K words) is processed by the five-phase pipeline into $M{=}119$ rules after verification. To prevent reward hacking through trivial pass-through, we automatically set $\mathrm{VFR}=0$ for any rewrite whose character similarity to the input exceeds $98\%$.

\paragraph{HumanEval~$\times$~PEP~8.}
HumanEval \citep{chen2021codex} provides $164$ function signatures and docstrings to be completed. PEP~8 ($\sim$8K words) is processed into $127$ atomic rules covering indentation, naming, whitespace, line wrapping, imports, expression formatting, and other style dimensions. For Pass@1, we prepend the prompt as setup code before executing the body, following the official HumanEval protocol (which handles cases such as HumanEval/38, where helper functions are defined in the prompt). The pylint evaluation uses the JSON2 output format with the default ruleset, with \texttt{missing-final-newline} and \texttt{trailing-newlines} disabled.

\paragraph{RuleArena-NBA.}
The NBA subset comprises $156$ problems stratified across three complexity levels: L0 ($61$ problems, single transaction), L1 ($69$ problems, two interacting transactions), and L2 ($26$ problems, three or more interacting transactions with sequential reasoning). The NBA Collective Bargaining Agreement ($\sim$60K words) is processed into $195$ atomic rules covering salary caps, contract limits, exception types
(BAE/NTMLE/MLE/supermax), free-agency mechanics, and trade
restrictions. For brevity, we use the following CBA abbreviations
throughout the paper: \textbf{MLE} (Mid-Level Exception),
\textbf{NTMLE} (Non-Taxpayer Mid-Level Exception, the higher MLE
variant available to non-tax teams), \textbf{BAE} (Bi-Annual
Exception), \textbf{TPE} (Traded Player Exception), and
\textbf{supermax} (Designated Veteran Player Extension).

\section{Evaluation Protocols}
\label{app:metrics}

\paragraph{NPOV scoring.}
The four 1--5 auxiliary scores are: \emph{Remediation} (how completely the rewrite resolves the flagged issue), \emph{Preservation} (how faithfully the rewrite preserves the original factual content), \emph{Tone} (whether the rewrite reads as encyclopaedic prose), and \emph{Fluency} (grammatical and stylistic naturalness). Both VFR and the auxiliary scores are produced by an LLM evaluator; the full evaluator prompt is in App.~\ref{app:prompts:eval-npov}.

\paragraph{Code metrics.}
Pass@1 is obtained from the official HumanEval test suite. Pylint operates over $[-10, 10]$. Radon reports the average cyclomatic complexity per function; Bandit reports the number of security issues flagged. Radon and Bandit are reported as secondary indicators only.

\paragraph{RuleArena-NBA strict accuracy.}
A problem is strict-correct only if the True/False answer, the identified offending transaction, and the identified responsible team are all correct. Partial credit is not awarded.

\section{Detailed Analysis for Rule Extraction Phases}\label{app:ablation-details-1}

\paragraph{NPOV results.}
Panel (a) of Table~\ref{tab:ablation1} reports the per-phase costs
on NPOV. Phase~3 (operationalization) contributes the most: removing
it costs $4.6$ VFR points ($68.2 \to 63.6$), the largest drop among
all variants. Without this phase, vague triggers such as ``when
appropriate'' and aspirational actions such as ``improve the tone''
remain in the rule set, and the matched rules, while still topically
correct, do not give the executor operations that can be carried out
without further interpretation. The drop suggests that downstream performance depends not only on which rules are selected, but also on whether the selected rules are concrete enough to act on, and that operationalization is what
turns otherwise plausible extracted prose into executable guidance.
Phases~2 and~5 are also clearly important, with costs of $3.7$ points
each ($68.2 \to 64.5$). Removing Phase~5 allows unverified rules whose source spans cannot be
matched in the policy to remain in the rule set; the matcher still selects a similar number
of rules per query ($24.8$ \#picks), but these unsupported rules may crowd out useful ones in the matched set
and can introduce instructions not grounded in the source document. Removing Phase~2 (atomic decomposition) leaves compound
rules intact, forcing the executor to disentangle
multiple prescribed actions per item and raising the chance that one
action is applied while another is dropped. We note that removing
Phase~2 \emph{reduces} \#picks ($22.1 \to 19.6$): without
decomposition the rule set contains fewer, coarser items, so the
matcher returns fewer matches, but each bundles several actions and is
harder to execute faithfully, so the cost reflects degraded
actionability per matched rule rather than an inflated matched set.
These two phases target failure modes distinct from Phase~3, namely
hallucination and bundling rather than vagueness, and their costs
confirm that all three are load-bearing components rather than
incremental refinements. Phases~4 (deduplication) and~1 (span
detection) have the smallest and equal effect on NPOV, each costing
$2.8$ points ($68.2 \to 65.4$). Without Phase~4, near-duplicate rules
inflate the matched set (\#picks rises from $22.1$ to $29.7$), but the
additional rules carry little new information, so the effect appears
to be mainly increased prompt load rather than a large change in the
available information. The small effect of Phase~1 we attribute to the
fact that the NPOV policy is densely normative, leaving relatively
little expository material for Phase~1 to filter.

\paragraph{RuleArena-NBA results.}
Panel (b) of Table~\ref{tab:ablation1} reports the same ablation on
a substantially different setting: a longer, more procedural source
document and a smaller open-weight executor. The dominant phase is
shared with NPOV. Phase~3 (operationalization) is again the single
largest contributor, with a drop of $6.7$ strict-accuracy points
($35.9 \to 29.2$). A similar pattern appears in NBA: without
operationalization, the rule set retains vague triggers and
under-specified actions, which on NBA translate into conditions such
as ``when the contract is sufficiently large'' or actions such as
``apply the relevant exception''. These survive into the matched set
but cannot be turned into a concrete legality judgment over specific
transactions. \#picks rises sharply (from $14.0$ to $34.0$),
indicating that vague conditions make the matcher less discriminative,
so the executor is simultaneously given more rules and less actionable
ones.

Phases~5, 4, and~2 form a second tier, with costs of $5.7$, $4.6$,
and $3.6$ points respectively. The internal ordering differs from
NPOV: Phase~4 (deduplication) is comparatively more costly on NBA, and
\#picks rises from $14.0$ to $22.0$ when it is removed. We attribute
this to the structure of the CBA, which restates similar provisions
across exception types (e.g., parallel rules for MLE, BAE, and NTMLE),
so near-duplicates are more prevalent than in NPOV and crowd the
matched set more aggressively when not collapsed. Phase~5
(verification) is also more costly on NBA than on NPOV, consistent
with a longer source document offering more opportunity for ungrounded
rules to survive without verification. Phase~2 (atomic decomposition)
is slightly less costly on NBA than on NPOV, likely because CBA
provisions are already drafted in a more itemized style, leaving fewer
compound rules to split. Phase~1 (span detection) is again the
smallest contributor ($-1.5$), even though the CBA contains more
non-normative content than the NPOV policy; we suspect the gain from
filtering is partially offset by the matcher's own ability to reject
non-applicable rules at inference time, and leave a finer-grained
analysis to future work.

\paragraph{Cross-Domain Consistency of Phase Contributions}
Across both panels of Table~\ref{tab:ablation1}, two contributions are
consistent despite substantial differences in source-document length,
document style, and executor type. Phase~3 (operationalization) is the
largest contributor in both settings, with drops of $4.6$ points on
NPOV ($68.2 \to 63.6$) and $6.7$ points on NBA ($35.9 \to 29.2$), and
in both cases it leads the next-most-costly phase by a clear margin.
At the other end, Phase~1 (span detection) is never the dominant
factor, costing $2.8$ points on NPOV and $1.5$ on NBA. The ranking of
the middle phases, however, is \emph{not} stable: on NPOV the second
tier is Phases~2 and~5 (tied at $3.7$), with Phases~1 and~4 below them
(tied at $2.8$), whereas on NBA the order is Phase~5 ($5.7$) $>$
Phase~4 ($4.6$) $>$ Phase~2 ($3.6$) $>$ Phase~1 ($1.5$). The fact that
the primacy of operationalization holds under a $4\times$ difference in
document length ($\sim$15K vs.\ $\sim$60K words) and across two
distinct executors suggests that this contribution reflects a
structural property of the extraction pipeline, while the reordering
of the middle phases reflects properties of the source document
itself, as detailed below.

\paragraph{Operationalization as the Dominant Bottleneck}
Phase~3 separates itself from the other phases by a clear margin on both domains. On NPOV, removing it incurs a $4.6$-point drop, which is $0.9$ points larger than the next-most-costly variants (Phases~2 and~5 at $3.7$); on NBA, the same ablation incurs a $6.7$-point drop, $1.0$ point larger than Phase~5 ($5.7$). A consistent secondary signal accompanies this drop: \#picks rises when Phase~3 is removed, from $22.1$ to $28.4$ on NPOV and, far more sharply, from $14.0$ to $34.0$ on NBA. The conjunction of a larger matched set and a lower downstream metric indicates that the failure mode is not under-retrieval but under-specification: vague conditions degrade the matcher's discriminativeness while simultaneously failing to provide actionable guidance to the executor, so additional rules translate into prompt bloat rather than useful instruction.

\paragraph{Domain-Sensitive Reordering Within the Second Tier}
Apart from the consistently dominant role of Phase~3 and the consistently minor role of Phase~1, the relative costs of the middle phases shift in interpretable ways between the two domains. Phase~4 (deduplication) is comparatively more costly on NBA ($4.6$ points, with \#picks rising from $14.0$ to $22.0$) than on NPOV ($2.8$ points, with \#picks rising from $22.1$ to $29.7$), consistent with the parallel-provision structure of the CBA, where similar clauses recur across exception types. Phase~5 (verification) is likewise more costly on NBA ($5.7$ vs.\ $3.7$ points), consistent with a longer source document leaving more room for ungrounded rules to survive. Conversely, Phase~2 (atomic decomposition) is less costly on NBA ($3.6$ points) than on NPOV ($3.7$ points), consistent with the more itemized drafting style of regulatory text, which leaves fewer compound rules to split. These shifts rearrange the middle phases but do not alter the position of Phase~3 as the largest contributor or Phase~1 as the smallest in either domain.

\paragraph{Bounded Absolute Gains on the Composition-Limited Domain}
The absolute spread of per-phase costs is comparable across domains ($4.6$ points on NPOV, $6.7$ points on NBA), but the ceilings differ substantially: the full pipeline reaches $68.2$ VFR on NPOV against $35.9$ strict accuracy on NBA. As a fraction of the full-pipeline metric, the cost of removing Phase~3 is therefore $6.7\%$ on NPOV but $18.7\%$ on NBA, meaning the extraction pipeline accounts for a substantially larger share of attainable performance on NBA in relative terms, even though the headroom in absolute points is broadly similar. This pattern is consistent with the broader finding in §\ref{sec:exp:nba} that NBA is bottlenecked by multi-step rule composition rather than rule selection alone: a clean, operationalized rule set is necessary to recover the gains that are recoverable, but the absolute level remains constrained by reasoning demands that no extraction phase can address.

\section{Detailed Analysis for Disentangling Retrieval Unit from Retrieval Relation}\label{app:ablation-details-2}
\methodname changes two aspects of standard RAG at once: the retrieval
\emph{unit} (text chunk \(\to\) atomic rule) and the retrieval \emph{relation}
(BGE similarity \(\to\) LLM pairwise applicability). The main results in
Table~\ref{tab:wiki-main} only compare the two corner cells (Chunks~+~Similarity
vs.\ Rules~+~Applicability), which conflates these two design choices. To
separate them, we run a $2\times2$ factorial ablation on NPOV that fills in
the two off-diagonal cells. \textsc{Rules~+~Similarity} applies BGE over the
same 119 atomic rules used by \methodname (embedding the concatenation of
\texttt{name}, \texttt{condition}, and \texttt{action}), with top-$k$ chosen
to approximately match the average number of picks of the applicability-based
matcher. \textsc{Chunks~+~Applicability} applies the same pairwise
applicability judge over the NPOV chunks. All four cells share the same
Qwen3-30B-A3B-Instruct-2507 executor and the same evaluation protocol.

Both axes contribute to the overall improvement, but the relation contributes
more. With the unit fixed at chunks, switching from similarity to applicability
raises VFR by 6.5 points ($53.3 \to 59.8$); with the unit fixed at rules, the
same switch raises VFR by a larger 10.3 points ($54.2 \to 64.5$). With the
relation fixed at similarity, switching from chunks to rules raises VFR by
only 0.9 points ($53.3 \to 54.2$); with the relation fixed at applicability,
the same switch raises VFR by 4.7 points ($59.8 \to 64.5$). On both axes,
therefore, the larger of the two single-factor lifts is observed when the
\emph{other} factor is already aligned with the downstream task, indicating
that the two design choices reinforce rather than substitute for each other.

This is confirmed by an additivity check at the diagonal
(\textsc{Rules~+~Applicability}, $64.5$): adding the two single-axis lifts
to the \textsc{Chunks~+~Similarity} cell gives a linear prediction of
$53.3 + 6.5 + 0.9 = 60.7$, which underestimates the observed $64.5$ by
$3.8$ points. The joint move from \textsc{Chunks~+~Similarity} to
\textsc{Rules~+~Applicability} therefore yields a modest super-additive gain
on top of the two main effects.

Two further patterns are worth noting. First, \textsc{Chunks~+~Applicability}
alone ($59.8$) already closes most of the gap between \textsc{Chunks~+~Similarity}
($53.3$) and the full method ($64.5$), suggesting that on NPOV the applicability
relation is the dominant single factor and that rule structuring contributes a
smaller, complementary refinement on top of it. Second, even when operating
over the same atomic-rule corpus, \textsc{Rules~+~Similarity} trails
\textsc{Rules~+~Applicability} by 10.3 points ($54.2$ vs.\ $64.5$). This rules
out the possibility that the gain is an artifact of feeding the executor a
cleaner retrieval unit, and confirms that the relation itself is decisive.
Together, these results support the central claim of the paper: the benefit
of \methodname on NPOV comes primarily from replacing similarity with
applicability, with rule structuring providing a smaller complementary
improvement that compounds with the relation change.

% =============================================================================
\section{Discussion}
\label{sec:discussion}

\subsection{When Does \methodname Help?}

The cross-domain results suggest three conditions under which the framework is most beneficial. The first is a high abstraction gap, where the inputs are concrete, and the applicable rules are stated abstractly, so that the two share little semantical similarity. NPOV violations evaluated against the NPOV policy are the clearest case. The second is strong conditionality, where each rule carries a sharp fire-when clause and applicability therefore becomes a well-defined binary question that an LLM judge can answer reliably; the CBA is structured in this way. The third is high distractor density, where the document contains many rules that look superficially similar to the input but are not the applicable ones, such as regular max-salary rules versus supermax extensions. Embedding similarity is most likely to fail within such regions.

\subsection{When Does It Not Help?}

Three settings limit the benefit. When rules and inputs share semantical similarity, as with PEP~8 style rules and Python code, a strong similarity retriever already retrieves the relevant context, and applicability matching adds little. When the total number of rules is small enough that an All Rules baseline fits comfortably in the context, retrieval ceases to be the bottleneck. And when the bottleneck is multi-step rule composition rather than rule selection, an improved retriever does not help; the L2 slice of RuleArena-NBA is precisely such a setting.

\subsection{What Are We Actually Contributing?}

We wish to be clear about what \methodname does and does not claim. The contribution is not a claim that LLMs are superior to dense retrievers in general, nor that pairwise judgment by an LLM is a universally better reranking step. The contribution is narrower: for documents whose useful units are conditional rules, retrieval should optimize condition satisfaction rather than semantical similarity, and this reframing is most consequential precisely where the two relations diverge. The experiments are designed to support that narrow claim by showing both where it holds (NPOV) and where it does not (PEP~8).

% =====================================================================
% Appendix F: Prompts  (v3 -- robust verbatim version)
=====================================================================

% ---------------------------------------------------------------------
% Prompt-box definition  (paste this in your main preamble OR keep here)
% ---------------------------------------------------------------------
\newtcolorbox{promptbox}[1][]{
    enhanced,
    breakable,
    colback=gray!5,
    colframe=gray!55,
    boxrule=0.4pt,
    arc=2pt,
    left=4pt, right=4pt, top=3pt, bottom=3pt,
    #1
}

% =====================================================================
% Appendix F: Prompts  (FINAL)
% TAG (Task-Aligned Retrieval)
%
% Required preamble packages (add to MAIN paper if not already present):
%   \usepackage{xcolor}
%   \usepackage{amsmath}
%   \usepackage{tcolorbox}
%   \tcbuselibrary{breakable,skins}
%   \usepackage{fvextra}   % NOT fancyvrb; fvextra supports breaklines
%
% Compile order: pdflatex twice for cross-references.
% =====================================================================

% ---------------------------------------------------------------------
% Prompt-box environment
% ---------------------------------------------------------------------
% \newtcolorbox{promptbox}[1][]{
%     enhanced,
%     breakable,
%     colback=gray!5,
%     colframe=gray!55,
%     boxrule=0.4pt,
%     arc=2pt,
%     left=4pt, right=4pt, top=3pt, bottom=3pt,
%     #1
% }

% =====================================================================
\section{Prompts}
\label{app:prompts}

This appendix lists, in full, every prompt used by TAG and by the
LLM-based evaluators reported in the main paper. All prompts are
reproduced verbatim from our experimental code, with runtime-substituted
fields written as \verb|{slot_name}|. Unless otherwise stated, every
LLM call uses temperature $0$, top-$p=1.0$, and a maximum output
length of \verb|4096| tokens. System and user messages are shown
separately wherever the underlying API distinguishes them.

We organise this appendix to mirror the three-stage pipeline of \S3.
\S\ref{app:prompts:extraction} covers offline rule extraction
(Stage~1); \S\ref{app:prompts:matcher} covers pairwise applicability
matching (Stage~2); \S\ref{app:prompts:executor} covers rule-guided
execution (Stage~3). \S\ref{app:prompts:eval} gives the LLM-as-judge
evaluation prompts, and \S\ref{app:prompts:rerank-baseline} gives the
controlled relevance-reranker baseline used to isolate the contribution
of the applicability framing.

% =====================================================================
\subsection{Stage 1: Rule Extraction Prompts}
\label{app:prompts:extraction}

The five-phase extraction pipeline converts an instructional document
into a set of structured condition--action rules. Phases~1--4 are
implemented as LLM calls with fixed JSON output schemas;
\textbf{Phase~5 is implemented programmatically as a verification step,
not as an additional LLM call.} The phases address distinct extraction
failure modes, identified in \S4.7: missed normative content
(Phase~1), bundled obligations (Phase~2), unjudgeable or inexecutable
rules (Phase~3), redundancy and conflict (Phase~4), and ungrounded or
uncovered rules (Phase~5).

% ---------------------------------------------------------------------
\subsubsection{Phase 1: Normative Span Detection}
\label{app:prompts:phase1}

\noindent\textbf{Purpose.}
Phase~1 identifies source spans that express prescriptive,
proscriptive, permissive, or conditional guidance. The objective is
high recall over normative content while preserving verbatim
traceability to the source document.

\medskip
\noindent\textbf{System message.}
\begin{promptbox}
\begin{Verbatim}[fontsize=\footnotesize,breaklines=true,breakanywhere=true]
You are an expert annotator for instructional and policy documents.

Your task is to identify normative spans: text segments that prescribe,
prohibit, recommend, permit, or conditionally require an action. A span
is normative if it tells an actor what should, must, may, should not, or
must not be done, or if it defines a condition under which such an
action applies.

Extract spans conservatively and preserve traceability to the source
document. Do not paraphrase source text.

Domain:
{domain}
\end{Verbatim}
\end{promptbox}

\medskip
\noindent\textbf{User message.}
\begin{promptbox}
\begin{Verbatim}[fontsize=\footnotesize,breaklines=true,breakanywhere=true]
Task:
Extract all normative spans from the document section below.

A normative span includes:
- requirements, obligations, or mandatory instructions;
- prohibitions or discouraged practices;
- recommendations, preferences, or best practices;
- permissions or exceptions that affect whether a rule applies;
- conditional statements that specify when an action should or should
  not be taken.

Exclude:
- purely descriptive background;
- historical or motivational prose;
- examples that do not themselves state a rule;
- definitions that do not constrain behavior, unless the definition is
  necessary to determine when a rule applies.

Document section:
"""
{doc}
"""

Output a JSON array. Each item must have the following fields:
{
  "span_id": "S-{start_id:03d}",
  "text": "<verbatim span copied exactly from the document>",
  "normative_type": "requirement" | "prohibition" | "recommendation"
                   | "permission" | "exception" | "conditional",
  "context_summary": "<one concise sentence describing the local topic>"
}

Requirements:
1. The "text" field must be an exact substring of the document section.
2. Do not rewrite, normalize, or merge non-contiguous text.
3. Prefer the shortest span that preserves the complete normative
   meaning.
4. Include all normative spans, including weak recommendations such as
   "prefer", "avoid", "normally", "should", and "recommended".
5. Return only the JSON array. Do not include markdown fences or
   explanations.
\end{Verbatim}
\end{promptbox}

% ---------------------------------------------------------------------
\subsubsection{Phase 2: Atomic Decomposition}
\label{app:prompts:phase2}

\noindent\textbf{Purpose.}
Phase~2 decomposes bundled normative spans into atomic units. An
atomic unit contains at most one primary subject, one applicability
context, and one prescribed action or constraint. This addresses the
\emph{bundling} failure mode identified in the ablation of \S4.7:
without decomposition, the executor must disentangle multiple
prescribed actions per matched item, increasing the chance that one
action is applied while another is dropped.

\medskip
\noindent\textbf{System message.}
\begin{promptbox}
\begin{Verbatim}[fontsize=\footnotesize,breaklines=true,breakanywhere=true]
You are an expert at decomposing normative language into atomic rule
units.

An atomic unit contains one primary subject, one applicability context,
and one prescribed action or constraint. The goal is to split bundled
spans into units that can later be converted into independently
checkable condition-action rules.

Preserve source traceability. Do not introduce new obligations that are
not entailed by the source span.

Domain:
{domain}
\end{Verbatim}
\end{promptbox}

\medskip
\noindent\textbf{User message.}
\begin{promptbox}
\begin{Verbatim}[fontsize=\footnotesize,breaklines=true,breakanywhere=true]
Task:
Decompose the following normative spans into atomic units.

Input spans:
{spans}

Decomposition rules:
1. Split a span when it contains multiple independently enforceable
   actions.
2. Split a span when it contains alternative conditions, such as
   "if A or B".
3. Split a span when different subjects are governed by different
   actions.
4. Do not split conjunctive conditions that must hold together, such
   as "if A and B".
5. Do not split exception clauses away from the rule they qualify.
6. Do not create a unit unless it is entailed by the original span.
7. Preserve the original wording as much as possible. If minor
   rewriting is needed to make the unit self-contained, keep the
   meaning faithful.

Output a JSON array. Each item must have the following fields:
{
  "atomic_id": "A-001",
  "source_span_id": "S-XXX",
  "text": "<self-contained atomic normative unit>",
  "original_text": "<verbatim source span text>",
  "was_split": true | false,
  "split_rationale": "<short reason if was_split is true, else null>"
}

Requirements:
- Each atomic unit should express at most one condition-action relation.
- The "original_text" field must be copied from the corresponding input
  span.
- Return only the JSON array. Do not include markdown fences or
  explanations.
\end{Verbatim}
\end{promptbox}

% ---------------------------------------------------------------------
\subsubsection{Phase 3: Operationalisation}
\label{app:prompts:phase3}

\noindent\textbf{Purpose.}
Phase~3 converts each atomic unit into an operational rule with a
\emph{judgeable} condition and an \emph{executable} action. The
ablation in Table~4 identifies this phase as the largest single
contributor to downstream performance, with removal costing $-4.6$
VFR on Wikipedia NPOV and $-6.7$ strict-accuracy points on
RuleArena-NBA. The dominance of this phase reflects the fact that
vague triggers and aspirational actions simultaneously degrade matcher
selectivity and reduce the actionability of the rules passed to the
executor.

\medskip
\noindent\textbf{System message.}
\begin{promptbox}
\begin{Verbatim}[fontsize=\footnotesize,breaklines=true,breakanywhere=true]
You are an expert at converting atomic normative statements into
structured, operational rules for instruction-following systems.

Each rule must have:
1. a judgeable condition: a state of the task input that can be checked
   by inspecting the input;
2. an executable action: a concrete operation that a downstream executor
   can perform;
3. a faithful source: the original text supporting the rule;
4. concise tags for grouping related rules.

Avoid vague, aspirational, or purely topical rules. The output must be
suitable for pairwise applicability matching and downstream execution.

Domain:
{domain}
\end{Verbatim}
\end{promptbox}

\medskip
\noindent\textbf{User message.}
\begin{promptbox}
\begin{Verbatim}[fontsize=\footnotesize,breaklines=true,breakanywhere=true]
Task:
Convert each atomic normative unit into an operational rule.

Input atomic units:
{atomics}

For each unit, produce one structured rule with the following fields:
{
  "rule_id": "R-001",
  "source_atomic_id": "A-XXX",
  "rule_name": "<3-8 word distinctive rule name>",
  "condition": "<detectable condition under which the rule applies>",
  "action": "<concrete operation to perform when the condition holds>",
  "source_text": "<verbatim source text supporting the rule>",
  "category_tags": ["<tag1>", "<tag2>"]
}

Quality requirements:
1. The condition must be checkable from a task input without external
   knowledge unless the task itself provides that knowledge.
2. The condition must not be merely topical. It should specify when the
   rule applies.
3. The action must specify what the executor should do, not merely state
   a goal.
4. Avoid vague actions such as "improve", "ensure quality",
   "be appropriate", or "make better" unless they are rewritten into
   concrete operations.
5. The source_text must be copied from the corresponding atomic unit's
   original source text.
6. Use 1-3 concise category_tags that capture the rule family.
7. Do not invent rules not supported by the atomic unit.

Examples of operationalisation:

(Wikipedia NPOV)
- Weak condition:  "The text is biased."
- Better condition: "The sentence states an evaluative characterization
                     in the author's voice without attribution."
- Weak action:     "Make the text neutral."
- Better action:   "Replace the evaluative characterization with a
                     neutral descriptive term or attribute it to a named
                     source if such a source is present."

(PEP 8)
- Weak condition:   "The code has import problems."
- Better condition: "A wildcard import (from <module> import *) is used."
- Weak action:      "Improve the imports."
- Better action:    "Replace the wildcard import with explicit imports
                     that list each imported name."

(NBA CBA)
- Weak condition:   "The transaction involves apron rules."
- Better condition: "A team attempts a transaction listed in the
                     Transaction Restrictions Table, and the transaction
                     has a defined Applicable Apron Level."
- Weak action:      "Check whether the transaction is allowed."
- Better action:    "Compute the team's salary immediately after the
                     transaction and flag a violation if it exceeds the
                     corresponding Applicable Apron Level."

Return only the JSON array. Do not include markdown fences or
explanations.
\end{Verbatim}
\end{promptbox}

% ---------------------------------------------------------------------
\subsubsection{Phase 4: Deduplication and Conflict Resolution}
\label{app:prompts:phase4}

\noindent\textbf{Purpose.}
Phase~4 compares rules that share at least one category tag and
identifies duplicate, subsuming, overlapping, or conflicting pairs.
Tag-restricted comparison avoids $O(M^2)$ scaling, while the resulting
labels support both automatic merging (for duplicates) and downstream
diagnostics (for conflicts).

\medskip
\noindent\textbf{System message.}
\begin{promptbox}
\begin{Verbatim}[fontsize=\footnotesize,breaklines=true,breakanywhere=true]
You are a rule-set quality auditor.

Your task is to compare candidate operational rules and identify
redundancy, overlap, and conflicts. The goal is to improve the rule set
while preserving coverage and traceability.

Classify only meaningful relationships. If two rules are independent,
omit them from the output.

Domain:
{domain}
\end{Verbatim}
\end{promptbox}

\medskip
\noindent\textbf{User message.}
\begin{promptbox}
\begin{Verbatim}[fontsize=\footnotesize,breaklines=true,breakanywhere=true]
Task:
Classify the relationship between each pair of rules below.

Rule pairs:
{rules}

Relationship labels:
- "duplicate": the two rules have substantially the same condition and
  prescribe substantially the same action. They should be merged or one
  should be removed.
- "subsumption": one rule is a more specific case of the other. Both
  may be kept if the narrower rule adds useful boundary information.
- "overlap": the rules share part of their applicability space but
  neither fully subsumes the other.
- "conflict": the rules can apply to the same input but prescribe
  incompatible actions.
- "independent": the rules govern different situations or actions.

Output only non-independent pairs.

For each non-independent pair, output:
{
  "rule_i": "R-XXX",
  "rule_j": "R-YYY",
  "relationship": "duplicate" | "subsumption" | "overlap" | "conflict",
  "preferred_action": "merge" | "keep_both" | "manual_review",
  "explanation": "<one concise sentence>"
}

Decision guidelines:
1. Use "duplicate" only when both condition and action are effectively
   the same.
2. Use "subsumption" when one rule is clearly more general and the other
   is a special case.
3. Use "overlap" when the rules intersect but neither rule contains the
   other.
4. Use "conflict" only when following both rules would lead to
   incompatible executor behavior.
5. Do not report weak topical similarity.

If all pairs are independent, return [].

Return only the JSON array. Do not include markdown fences or
explanations.
\end{Verbatim}
\end{promptbox}

% ---------------------------------------------------------------------
\subsubsection{Phase 5: Verification}
\label{app:prompts:phase5}

\noindent\textbf{Purpose.}
Phase~5 verifies the extracted rule set using \emph{programmatic}
checks rather than an additional LLM call. It reports three
diagnostics -- Faithfulness, Coverage, and Independence -- and acts on
each according to the policy below. Faithfulness failures result in
\emph{rule removal}; Coverage failures \emph{trigger Phase~3
re-extraction} over the uncovered spans; Independence is reported as a
\emph{diagnostic} and does not by itself cause rule removal.

\medskip
\noindent\textbf{Faithfulness.}
For each rule, we verify whether its \verb|source_text| can be matched
to the original document. Exact substring matches are accepted
directly. If no exact match is found, we compute a character-level
fuzzy match using
\texttt{difflib.SequenceMatcher(\detokenize{None}, src, w).ratio()}
over sliding windows $w$ of the document, with stride $50$ characters
and window length $|\text{src}|+50$. A rule is counted as faithful
when the best match score exceeds
$\tau_{\text{faith}} = 0.85$. Rules whose source cannot be matched
above $\tau_{\text{faith}}$ are discarded.

\medskip
\noindent\textbf{Coverage.}
For each normative span $s_j$ extracted in Phase~1, we check whether at
least one surviving operational rule has a \verb|source_text| that
overlaps $s_j$ at the character level by at least $50\%$. Uncovered
spans are sent back through Phase~3 for re-extraction; spans still
uncovered after re-extraction are logged and excluded from the final
rule set.

\medskip
\noindent\textbf{Independence.}
We compute the ratio of unique rule names to total rules after
deduplication. The ratio is reported as a diagnostic and is not used
to drop rules; rules that survive Phase~4 but share a name with another
surviving rule are flagged for manual inspection only.

% =====================================================================
\subsection{Stage 2: Pairwise Applicability Matching Prompts}
\label{app:prompts:matcher}

For methods using applicability-based retrieval, we evaluate each
input--rule pair independently. The matcher is asked to decide whether
the input satisfies the rule's condition, rather than whether the rule
is topically similar or generally useful.

\medskip
\noindent\textbf{Design note.}
In both prompts below, only the rule's \verb|rule_id|,
\verb|rule_name|, \verb|condition|, and tag-level metadata are exposed
to the matcher; the \verb|action| field is deliberately withheld.
Exposing the action biases the judge toward rules with appealing
actions even when the condition does not hold, conflating
applicability with action desirability. This withholding is the
central distinction between the applicability matcher used by TAG and
the controlled relevance reranker reported in
\S\ref{app:prompts:rerank-baseline}.

A single \emph{general} applicability matcher is used for Wikipedia
NPOV and HumanEval + PEP~8 (\S\ref{app:prompts:matcher-general}). The
RuleArena-NBA matcher uses a domain-specific schema with explicit
boundary-condition handling (\S\ref{app:prompts:matcher-nba}),
reflecting the structure of the CBA, where many rules differ only by
exception type, transaction type, or contract class.

% ---------------------------------------------------------------------
\subsubsection{General Applicability Matcher}
\label{app:prompts:matcher-general}

\noindent\textbf{Purpose.}
For each pair $(x, r_i)$, the matcher decides whether the input $x$
satisfies the condition of rule $r_i$. The decision is intentionally
formulated as condition satisfaction rather than topical relevance or
general usefulness.

\medskip
\noindent\textbf{Input.}
A task input $x$ and one candidate rule $r_i$. Only the rule's
\verb|rule_id|, \verb|rule_name|, \verb|condition|, and
\verb|category_tags| are shown to the matcher. The rule's
\verb|action| is withheld to prevent the judge from selecting rules
because their actions appear useful even when their conditions do not
hold.

\medskip
\noindent\textbf{Output.}
A JSON object with a binary verdict: \verb|{"verdict": "YES"}| or
\verb|{"verdict": "NO"}|.

\medskip
\noindent\textbf{System message.}
\begin{promptbox}
\begin{Verbatim}[fontsize=\footnotesize,breaklines=true,breakanywhere=true]
You are a strict applicability judge for instruction-following tasks.

Given one task input and one candidate rule extracted from an
instruction document, decide whether the rule is APPLICABLE to the
input. A rule is applicable if, and only if, the observable condition
of the rule is satisfied by the input.

Your decision should be based on condition satisfaction, not topical
similarity, general usefulness, or whether the rule sounds helpful.
Be conservative: if the condition is ambiguous, underspecified, or not
directly supported by the input, answer NO.

Output ONLY a single JSON object. Do not include reasoning,
explanations, markdown fences, or any other text.
\end{Verbatim}
\end{promptbox}

\medskip
\noindent\textbf{User message.}
\begin{promptbox}
\begin{Verbatim}[fontsize=\footnotesize,breaklines=true,breakanywhere=true]
Task input:
"""
{query}
"""

Candidate rule:
- rule_id: {rule_id}
- rule_name: {rule_name}
- condition: {condition}
- category_tags: {tags}

Question:
Does the task input satisfy the candidate rule's condition?

Decision criteria:
- Answer YES only if the condition clearly holds for this specific
  input.
- Answer NO if the rule is merely topically related but its condition
  is not satisfied.
- Answer NO if deciding applicability would require information not
  present in the input.
- Answer NO if applying the rule would require inventing facts,
  assumptions, sources, or hidden context.
- Borderline or uncertain cases default to NO.

Output format:
{"verdict": "YES"} or {"verdict": "NO"}
\end{Verbatim}
\end{promptbox}

% ---------------------------------------------------------------------
\subsubsection{RuleArena-NBA Matcher}
\label{app:prompts:matcher-nba}

The NBA Collective Bargaining Agreement contains many rules that
differ only by boundary conditions on a particular exception type,
transaction type, or contract class. The matcher prompt therefore
exposes an expanded rule schema (\verb|applies_when|, \verb|constraint|,
\verb|violation_check|, \verb|does_not_apply_to|) and adds explicit
decision criteria for distinguishing exception-specific from generic
regulatory rules.

\medskip
\noindent\textbf{System message.}
\begin{promptbox}
\begin{Verbatim}[fontsize=\footnotesize,breaklines=true,breakanywhere=true]
You are a strict applicability judge for NBA Collective Bargaining
Agreement (CBA) rules.

Given one transaction scenario and one candidate CBA rule, decide
whether the rule is applicable to evaluating the scenario. A rule is
applicable only if the scenario contains observable facts satisfying
the rule's non-trivial preconditions. Do not select a rule merely
because it is topically related to salary-cap reasoning.

Be conservative. If the scenario does not provide enough information
to verify the rule's preconditions, or if a veto condition applies,
answer false.

Output only a single JSON object.
\end{Verbatim}
\end{promptbox}

\medskip
\noindent\textbf{User message.}
\begin{promptbox}
\begin{Verbatim}[fontsize=\footnotesize,breaklines=true,breakanywhere=true]
Scenario:
"""
{scenario}
"""

Candidate rule:
- rule_id: {rule_id}
- rule_name: {rule_name}
- primary_tag: {primary_tag}
- applies_when: {applies_when}
- constraint: {constraint}
- violation_check: {violation_check}
- does_not_apply_to:
{does_not_apply_to}

Question:
Is this candidate rule applicable to evaluating the transaction
scenario?

Decision criteria:
1. Match the rule to the primary transaction type in the scenario,
   such as signing, trade, extension, renegotiation, qualifying offer,
   or sign-and-trade. Reject the rule if its transaction type does
   not match the scenario.
2. Answer true only when the scenario contains at least one
   substantive fact satisfying the rule's non-trivial precondition.
3. Do not treat generic procedural phrases, such as "evaluate",
   "check", "compute", or "test", as substantive preconditions.
4. If the rule concerns a specific exception, contract type, or
   mechanism, such as MLE, BAE, TPE, two-way contract, 10-day
   contract, rookie extension, or designated veteran extension, the
   scenario must mention that mechanism or clearly instantiate it.
5. If the rule governs a general regulatory constraint, such as cap,
   tax, apron, maximum salary, salary growth, or contract length, the
   scenario does not need to name the legal term explicitly, but it
   must contain the facts needed to evaluate the constraint.
6. If any item in does_not_apply_to matches the scenario, answer
   false.
7. When several closely related rules differ only by boundary
   conditions, answer true only for the rule whose boundary condition
   is specifically supported by the scenario.
8. Borderline or uncertain cases default to false.

Output exactly one JSON object:
{
  "applicable": true | false,
  "reason": "<one short sentence citing the decisive scenario fact or veto>"
}
\end{Verbatim}
\end{promptbox}

% =====================================================================
\subsection{Stage 3: Executor Prompts}
\label{app:prompts:executor}

The executor prompt is held fixed within each retrieval-unit type:
rule-mode prompts are used by TAG and by the All-Rules baseline, while
chunk-mode prompts are used by Standard RAG. Methods differ only in
the reference material supplied to the executor
(\verb|{reference_rules}| in the templates below), allowing the
comparison in the main tables to isolate the retrieval mechanism.

% ---------------------------------------------------------------------
\subsubsection{Wikipedia NPOV Rewriting Executor}
\label{app:prompts:exec-npov}

\noindent\textbf{Rule-mode system message.}
\begin{promptbox}
\begin{Verbatim}[fontsize=\footnotesize,breaklines=true,breakanywhere=true]
You are a Wikipedia editor revising a sentence for compliance with
Wikipedia's Neutral Point of View (NPOV) policy.

You will receive one input sentence and a list of candidate rewrite
actions extracted from the NPOV policy. Treat the provided actions as
the only policy guidance available for this task. Apply any action
whose required edit can be performed using only information already
present in the input sentence.

Do not introduce new facts, sources, citations, viewpoints, or external
context. If an action requires information not present in the input,
do not apply that action. Preserve the original topic and factual
content as much as possible while making the sentence more neutral.

Output exactly three lines in the following format:
Applied rules: <comma-separated rule IDs from the provided list, or NONE>
Reasoning: <one concise sentence explaining the concrete edit performed>
Rewrite: <the revised sentence on a single line>

If no provided action can be applied without adding external
information, write Applied rules: NONE and copy the input sentence
unchanged in the Rewrite line.
\end{Verbatim}
\end{promptbox}

\medskip
\noindent\textbf{Rule-mode user message.}
\begin{promptbox}
\begin{Verbatim}[fontsize=\footnotesize,breaklines=true,breakanywhere=true]
Task:
Revise the input sentence to better comply with Wikipedia's Neutral
Point of View policy by applying the provided rewrite actions.

Candidate rewrite actions:
{reference_rules}

Input sentence:
"""
{query}
"""

Instructions:
1. Apply only actions whose required edit can be performed using
   information already present in the input sentence.
2. Do not add new facts, sources, citations, viewpoints, or external
   context.
3. Preserve the original factual content and topic as much as possible.
4. If multiple actions can be applied, apply the subset that produces
   the most direct and minimal neutral rewrite.
5. If no provided action can be applied without adding external
   information, output NONE and leave the sentence unchanged.

Output exactly three lines:
Applied rules: <comma-separated rule IDs from the provided list, or NONE>
Reasoning: <one concise sentence explaining the concrete edit performed>
Rewrite: <the revised sentence on a single line>
\end{Verbatim}
\end{promptbox}

\medskip
\noindent\textbf{Chunk-mode system message.}
\begin{promptbox}
\begin{Verbatim}[fontsize=\footnotesize,breaklines=true,breakanywhere=true]
You are a Wikipedia editor revising a sentence for compliance with
Wikipedia's Neutral Point of View (NPOV) policy.

You will receive one input sentence and several excerpts from the NPOV
policy. Treat the provided excerpts as the only policy guidance
available for this task. Apply any concrete guidance from the excerpts
whose required edit can be performed using only information already
present in the input sentence.

Do not introduce new facts, sources, citations, viewpoints, or external
context. If the guidance requires information not present in the input,
do not apply it. Preserve the original topic and factual content as
much as possible while making the sentence more neutral.

Output exactly three lines in the following format:
Applied rules: POLICY or NONE
Reasoning: <one concise sentence explaining the concrete guidance applied>
Rewrite: <the revised sentence on a single line>

If no policy guidance can be applied without adding external
information, write Applied rules: NONE and copy the input sentence
unchanged in the Rewrite line.
\end{Verbatim}
\end{promptbox}

\medskip
\noindent\textbf{Chunk-mode user message.}
\begin{promptbox}
\begin{Verbatim}[fontsize=\footnotesize,breaklines=true,breakanywhere=true]
Task:
Revise the input sentence to better comply with Wikipedia's Neutral
Point of View policy by using the provided policy excerpts.

Policy excerpts:
{reference_rules}

Input sentence:
"""
{query}
"""

Instructions:
1. Apply only guidance from the excerpts whose required edit can be
   performed using information already present in the input sentence.
2. Do not add new facts, sources, citations, viewpoints, or external
   context.
3. Preserve the original factual content and topic as much as possible.
4. If multiple pieces of guidance can be applied, apply the subset that
   produces the most direct and minimal neutral rewrite.
5. If no provided guidance can be applied without adding external
   information, output NONE and leave the sentence unchanged.

Output exactly three lines:
Applied rules: POLICY or NONE
Reasoning: <one concise sentence explaining the concrete guidance applied>
Rewrite: <the revised sentence on a single line>
\end{Verbatim}
\end{promptbox}

% ---------------------------------------------------------------------
\subsubsection{HumanEval + PEP 8 Executor}
\label{app:prompts:exec-code}

The executor receives the original HumanEval function signature and
docstring as \verb|{prompt}|, and a set of style-guidance items as
\verb|{reference_rules}|. The same prompt is used for both rule-mode
and chunk-mode retrieval; only the form of \verb|{reference_rules}|
changes.

\medskip
\noindent\textbf{System message.}
\begin{promptbox}
\begin{Verbatim}[fontsize=\footnotesize,breaklines=true,breakanywhere=true]
You are a Python code generation system. Complete the given Python
function according to its signature and docstring.

You may also receive coding-style rules or reference guidance. Treat
the provided material as the only style guidance available for this
task. Apply any rule whose condition is relevant to the code you
write, but do not sacrifice functional correctness for style.

Output only plain Python source code. Do not use markdown fences. Do
not include explanations, tests, or extra text.
\end{Verbatim}
\end{promptbox}

\medskip
\noindent\textbf{User message.}
\begin{promptbox}
\begin{Verbatim}[fontsize=\footnotesize,breaklines=true,breakanywhere=true]
Task:
Complete the following Python function.

Function to complete:
{prompt}

Coding-style rules or reference guidance:
{reference_rules}

Instructions:
1. Output the complete function, including the original signature and
   docstring.
2. The function name and signature must match the prompt exactly.
3. Preserve the provided docstring as-is.
4. Functional correctness is the top priority.
5. Apply every provided coding-style rule whose condition is relevant
   to the code you produce.
6. Do not introduce unnecessary imports, comments, dead code, unused
   variables, or overly complex control flow.
7. Output only Python source code, with no markdown fences or
   explanations.
\end{Verbatim}
\end{promptbox}

% ---------------------------------------------------------------------
\subsubsection{RuleArena-NBA Executor}
\label{app:prompts:exec-nba}

\noindent\textbf{System message.}
\begin{promptbox}
\begin{Verbatim}[fontsize=\footnotesize,breaklines=true,breakanywhere=true]
You are an NBA salary-cap analyst. Determine whether the proposed
transactions comply with the NBA Collective Bargaining Agreement.

You will receive one transaction scenario and a set of CBA rules or
reference guidance selected by a retrieval method. Use only the
provided reference material and the facts stated in the scenario. Do
not invent missing contract terms, cap figures, exceptions, or team
circumstances.

Return a JSON object only. Do not include markdown fences,
explanations outside the JSON, or extra text.
\end{Verbatim}
\end{promptbox}

\medskip
\noindent\textbf{User message.}
\begin{promptbox}
\begin{Verbatim}[fontsize=\footnotesize,breaklines=true,breakanywhere=true]
Task:
Evaluate whether the proposed NBA transaction scenario is legal under
the Collective Bargaining Agreement.

Scenario:
"""
{scenario}
"""

CBA rules or reference guidance:
{reference_rules}

Instructions:
1. Determine whether the transaction is legal using the provided
   reference material and the facts stated in the scenario.
2. If the transaction is legal, set "answer" to false and set both
   "illegal_operation" and "problematic_team" to null.
3. If the transaction is illegal, set "answer" to true, identify the
   specific illegal operation, and identify the responsible team.
4. Do not rely on unstated facts, unstated exceptions, or external
   knowledge.
5. If the provided reference material is insufficient to establish
   illegality, prefer "answer": false.

Output exactly one JSON object:
{
  "answer": true | false,
  "illegal_operation": "<operation id or short description, or null>",
  "problematic_team": "<team name, or null>",
  "rationale": "<one short sentence>"
}
\end{Verbatim}
\end{promptbox}

% =====================================================================
\subsection{Evaluation Prompts}
\label{app:prompts:eval}

All NPOV metrics are produced by an LLM judge; HumanEval
\verb|Pass@1| and PEP~8 \verb|pylint| scores are computed by their
respective standard tools; and RuleArena-NBA strict accuracy is
computed programmatically from the executor's JSON output. The
LLM-judged prompt is given below.

% ---------------------------------------------------------------------
\subsubsection{NPOV Violation Fix Rate and Auxiliary Scores Judge}
\label{app:prompts:eval-npov}

A single LLM judge produces both the binary Violation Fix Rate (VFR)
label and the four 1--5 auxiliary scores (Remediation, Preservation,
Tone, Fluency) reported in Table~2. To prevent reward hacking via
pass-through rewrites, any candidate rewrite whose character similarity
to the original input exceeds $0.98$ is automatically assigned
$\text{VFR}=0$ before this prompt is invoked.

\medskip
\noindent\textbf{System message.}
\begin{promptbox}
\begin{Verbatim}[fontsize=\footnotesize,breaklines=true,breakanywhere=true]
You are a strict and fair Wikipedia Neutral Point of View (NPOV)
evaluation judge.

Your task is to evaluate whether a rewritten sentence fixes the
specific NPOV violation described for the original sentence, while
preserving the original factual content. Judge only the rewrite
quality; do not reward or penalize the retrieval method that produced
it.

Be strict about the target violation: a rewrite receives VFR=true only
if the specific violation described in the input is fully resolved.
Partial fixes, generic improvements, unsupported additions, or
rewrites that create a new comparable neutrality problem should
receive VFR=false.

Output only a single JSON object. Do not include markdown fences,
comments, or any extra text.
\end{Verbatim}
\end{promptbox}

\medskip
\noindent\textbf{User message.}
\begin{promptbox}
\begin{Verbatim}[fontsize=\footnotesize,breaklines=true,breakanywhere=true]
Task:
Evaluate a single attempted rewrite of a Wikipedia sentence.

Original sentence:
"""
{original}
"""

Specified NPOV violation:
"""
{violation}
"""

Rewrite to evaluate:
"""
{rewrite}
"""

Evaluation dimensions:

1. VFR (Violation Fix Rate), binary:
   - true: the rewrite fully resolves the specified NPOV violation.
   - false: the specified violation remains, is only partially
     addressed, or is replaced by a new neutrality problem of
     comparable severity.

2. Remediation, integer from 1 to 5:
   - 1: the specified violation is not addressed.
   - 2: minimal or mostly ineffective attempt.
   - 3: partial fix with a clear remaining issue.
   - 4: mostly addresses the violation with minor weakness.
   - 5: fully addresses the specified violation.

3. Preservation, integer from 1 to 5:
   - 1: key factual content is removed, contradicted, or fabricated.
   - 2: important facts are substantially altered or omitted.
   - 3: some factual content is changed or lost, main point remains.
   - 4: nearly all factual content preserved with only minor
     compression.
   - 5: all factual content relevant to the original is preserved.
   Penalize rewrites that fix neutrality by deleting essential facts.

4. Tone, integer from 1 to 5:
   - 1: clearly biased, promotional, judgmental, or inflammatory.
   - 2: noticeably non-neutral.
   - 3: mostly neutral but with some evaluative wording.
   - 4: neutral with minor stylistic weakness.
   - 5: fully neutral and encyclopedic in tone.

5. Fluency, integer from 1 to 5:
   - 1: ungrammatical or difficult to understand.
   - 2: awkward or unclear.
   - 3: readable but stilted.
   - 4: fluent with minor issues.
   - 5: polished, natural encyclopedic prose.

Additional judging rules:
- Do not give VFR=true merely because the rewrite is shorter or more
  generic.
- Do not give VFR=true if the rewrite removes the problematic language
  by dropping essential factual content.
- Do not reward unsupported attribution, added sources, added
  viewpoints, or factual claims not present in the original sentence.
- A near-copy of the original should receive VFR=false unless the
  specified violation is actually removed.
- Scores should be integers only.

Output exactly one JSON object:
{
  "VFR": true | false,
  "Rem":  <integer 1-5>,
  "Pres": <integer 1-5>,
  "Tone": <integer 1-5>,
  "Flu":  <integer 1-5>,
  "reason": "<one concise sentence explaining the main judgment>"
}
\end{Verbatim}
\end{promptbox}

% =====================================================================
\subsection{Controlled Baseline: LLM Relevance Reranker}
\label{app:prompts:rerank-baseline}

To isolate the contribution of the \emph{applicability} framing
specifically -- as opposed to merely substituting an LLM judge for a
dense retriever -- we additionally evaluate a controlled baseline that
uses the same pairwise LLM-judgment setup as TAG, but asks a standard
relevance question. The baseline differs from the applicability
matcher of \S\ref{app:prompts:matcher} in exactly two respects:
(i)~the predicate is \emph{relevance} rather than \emph{condition
satisfaction}, and (ii)~the rule's \verb|action| field is exposed,
matching the standard LLM-reranker setup of
\citet{nogueira2019passage} and \citet{glass2022rerank}.
%% TODO: change the two \citet keys above to whatever your bibfile
%% actually uses for Nogueira and Cho (2019) and Glass et al. (2022).
All other elements -- the rule set, the executor prompt, the model,
and the decoding settings -- are held fixed across the two conditions.

\medskip
\noindent\textbf{System message.}
\begin{promptbox}
\begin{Verbatim}[fontsize=\footnotesize,breaklines=true,breakanywhere=true]
You are a strict relevance judge for instruction-following tasks.

Given one task input and one candidate rule extracted from an
instruction document, decide whether the rule is RELEVANT to the
input. A rule is relevant if it is plausibly useful for handling the
input, regardless of whether its specific condition is satisfied.

Output ONLY a single JSON object. Do not include reasoning,
explanations, markdown fences, or any other text.
\end{Verbatim}
\end{promptbox}

\medskip
\noindent\textbf{User message.}
\begin{promptbox}
\begin{Verbatim}[fontsize=\footnotesize,breaklines=true,breakanywhere=true]
Task input:
"""
{query}
"""

Candidate rule:
- rule_id: {rule_id}
- rule_name: {rule_name}
- condition: {condition}
- action: {action}
- category_tags: {tags}

Question:
Is this candidate rule relevant to the task input?

Output format:
{"verdict": "YES"} or {"verdict": "NO"}
\end{Verbatim}
\end{promptbox}

\noindent The contrast between this prompt and the applicability
matcher in \S\ref{app:prompts:matcher} is the cleanest available
prompt-level test of whether the applicability framing -- rather than
LLM judgment alone -- accounts for TAG's gains.

% =====================================================================
% End of Appendix F
% =====================================================================

\input{tex/appendix_G.tex}

\end{document}

%% file: tex/appendix_G.tex
% =============================================================================
% Appendix G: Human Evaluation on Wikipedia NPOV
% =============================================================================

\section{Human Evaluation on Wikipedia NPOV}
\label{app:human_eval}

To validate the LLM-as-a-judge scores used in Section~4.3, we conduct a
small-scale human evaluation on a 40-query subset of the Wikipedia NPOV
evaluation set. Two annotators independently rate anonymized outputs from
three systems: All Rules (M1), Standard RAG top-20 (M2), and TAG (M3, ours).
We use this study as a trend-level validation rather than as a replacement
for the full automatic evaluation on all 107 cases.

\subsection{Protocol}
\label{app:human_eval:protocol}

Each query is shown with the original sentence, the editor-written violation
description, and three candidate rewrites. The rewrites are presented under
blinded labels A/B/C, and annotators are not told which method produced
which rewrite. Annotators rate each rewrite on four dimensions: VFR
(binary), Remediation, Preservation, and Tone. The three 1--5 scores follow
the definitions used by the LLM evaluator in Appendix~F.4.1.

The subset consists of Q001--Q040 from the 107-case NPOV evaluation set.
Since this subset is not a random sample, we report the LLM-judge scores on
the same 40 cases when comparing human and automatic evaluation.

\subsection{Annotator Recruitment and Consent}
\label{app:human_eval:annotators}

We recruited two university students as volunteer annotators, who were
compensated for their time. Both annotators were given written
instructions describing the Wikipedia NPOV rewriting task and the rating
criteria. For each item, they were shown the original sentence, the
violation description, and the anonymized system rewrites, and were asked
to rate whether each rewrite fixed the violation and to score remediation,
preservation, and tone (the rating dimensions detailed in
Appendix~\ref{app:human_eval:protocol}). Annotators were informed that
their ratings would be used for research purposes in aggregate form. We did
not collect any personal or sensitive information from the annotators beyond
their ratings.

\subsection{Human Scores}
\label{app:human_eval:scores}

Table~\ref{tab:appG_human_scores} reports the human evaluation results.
TAG obtains the highest human VFR and Remediation score, and the overall
ranking on the main metric is M3 $>$ M1 $>$ M2.

\begin{table}[h]
\centering
\small
\setlength{\tabcolsep}{4pt}
\begin{tabular}{lcccc}
\toprule
Method & VFR & Rem. & Pres. & Tone \\
\midrule
All Rules (M1) & 85.0 & 4.28 & 4.85 & 4.37 \\
Std RAG top-20 (M2) & 65.0 & 3.41 & 4.83 & 3.51 \\
TAG (M3, ours) & 92.4 & 4.47 & 4.88 & 4.06 \\
\bottomrule
\end{tabular}
\caption{Human evaluation on 40 Wikipedia NPOV cases. VFR is reported as a
percentage; Remediation, Preservation, and Tone are 1--5 scores averaged
over two annotators.}
\label{tab:appG_human_scores}
\end{table}

\FloatBarrier

\subsection{Comparison with the LLM Judge}
\label{app:human_eval:human_vs_llm}

Table~\ref{tab:appG_side_by_side} compares human scores with the LLM-judge
scores on the same 40-query subset. Human annotators assign higher absolute
VFR scores than the LLM judge, suggesting that the LLM judge is more
conservative on borderline rewrites. However, the method ranking is
preserved: TAG performs best, followed by All Rules and Standard RAG
top-20.

\begin{table}[h]
\centering
\small
\setlength{\tabcolsep}{5pt}
\begin{tabular}{lcc}
\toprule
Method & LLM VFR & Human VFR \\
\midrule
All Rules (M1) & 55.0 & 85.0 \\
Std RAG top-20 (M2) & 47.5 & 65.0 \\
TAG (M3, ours) & 70.0 & 92.4 \\
\bottomrule
\end{tabular}
\caption{LLM-judge and human VFR on the same 40-query subset. Human
annotators are more permissive in absolute scores, but preserve the same
ranking as the LLM judge.}
\label{tab:appG_side_by_side}
\end{table}

\FloatBarrier

\subsection{Agreement and Significance}
\label{app:human_eval:agreement}

Inter-annotator agreement is substantial on the main VFR metric
(Cohen's $\kappa=0.68$). Agreement is also substantial on Remediation and
Tone under quadratic-weighted $\kappa$, while Preservation has lower
chance-corrected agreement because most rewrites receive high Preservation
scores, leaving little variance.

On the 40-query subset, TAG outperforms Standard RAG top-20 on human VFR
by 27.5 percentage points. This difference is significant under an exact
McNemar test ($p=0.003$). TAG is not significantly different from All Rules,
which is consistent with the main-paper result that rule structuring already
captures a substantial portion of the available gain, while applicability
selection adds a smaller improvement.

\subsection{Subset Representativeness}
\label{app:human_eval:representativeness}

The 40-query subset is not a random sample of the full 107-case set. To
avoid comparing different sample sets, we compare human evaluation only
against the LLM-judge scores on the same 40 cases. Under the LLM judge, the
subset scores are 55.0\%, 47.5\%, and 70.0\% VFR for M1, M2, and M3,
respectively, preserving the same ranking as the full-set results in
Table~1. This suggests that the human evaluation supports the same
system-level trend observed in the main experiment, while the full-set
automatic evaluation remains the primary result.

\subsection{Summary}
\label{app:human_eval:summary}

The human evaluation provides a small-scale validation of the main NPOV
trend. Although human annotators are more permissive than the LLM judge in
absolute VFR, both evaluations rank the methods in the same order:
TAG $>$ All Rules $>$ Standard RAG top-20. We therefore treat the human
study as supporting evidence that TAG's improvement is not solely an
artifact of the LLM-based evaluator.